# Comparing GC and Field LMXBs in Elliptical Galaxies with deep *Chandra* and *Hubble* data

D.-W. Kim[1], G. Fabbiano[1], N. J. Brassington[1], T. Fragos[2], V. Kalogera[2], A. Zezas[1], A. Jordán[1,3], G. R. Sivakoff[4], A. Kundu[5], S. E. Zepf[5], L. Angelini[6], R. L. Davies[7], J. S. Gallagher[8], A. M. Juett[6], A. R. King[9], S. Pellegrini[10], C. L. Sarazin[4], G. Trinchieri[11]

[1] Harvard-Smithsonian Center for Astrophysics, 60 Garden St., Cambridge MA 02138; kim@cfa.harvard.edu, gfabbiano@cfa.harvard.edu, azezas@cfa.harvard.edu
[2] Northwestern University, Department of Physics and Astronomy, 2145 Sheridan Road, Evanston, IL 60208; vicky@northwestern.edu
[3] Departamento de Astronomía y Astrofísica, Pontificia Universidad Católica de Chile, Casilla 306, Santiago 22, Chile; ajordan@astro.puc.cl
[4] Department of Astronomy, University of Virginia, VA 22904; grs8g@virginia.edu, cls7i@virginia.edu
[5] Department of Physics and Astronomy, Michigan State University, East Lansing, MI 48824-2320; zepf@pa.msu.edu
[6] Laboratory for High Energy Astrophysics, NASA Goddard Space Flight Center, Code 660, Greenbelt, MD 20771; angelini@davide.gsfc.nasa.gov, Adrienne.M.Juett@nasa.gov
[7] Denys Wilkinson Building, University of Oxford, Keble Road, Oxford; rld@astro.ox.ac.uk
[8] Astronomy Department, University of Wisconsin, 475 North Charter Street, Madison, WI 53706; jsg@astro.wisc.edu
[9] University of Leicester, Leicester, LE1 7RH, UK; ark@star.le.ac.uk
[10] Dipartimento di Astronomia, Universita' di Bologna, Via Ranzani 1, 40127, Bologna, Italy; silvia.pellegrini@unibo.it
[11] INAF-Osservatorio Astronomico di Brera, via Brera 28, 20121 Milano, Italy; ginevra.trinchieri@brera.inaf.it

(July 07, 2009)

We present a statistical study of the low-mass X-ray binary (LMXB) populations of three nearby, old elliptical galaxies: NGC 3379, NGC 4278, and NGC 4697. With a cumulative ~1 Ms *Chandra* ACIS observing time, we detect 90-170 LMXBs within the $D_{25}$ ellipse of each galaxy. Cross-correlating *Chandra* X-ray sources and *HST* optical sources, we identify 75 globular cluster (GC) LMXBs and 112 field LMXBs with $L_X > 10^{36}$ erg s$^{-1}$ (detections of these populations are 90% complete down to luminosities in the range of 6 x $10^{36}$ – 1.5 x$10^{37}$ erg s$^{-1}$). At the higher luminosities explored with previous studies, the statistics of this sample are consistent with the properties of GC-LMXBs reported in the literature. In the low luminosity range allowed by our deeper data ($L_X < 5$ x $10^{37}$ erg s$^{-1}$), we find a significant relative lack of GC-LMXBs, when compared with field sources. Using the co-added sample from the three galaxies, we find that the incompleteness-corrected X-ray luminosity functions (XLFs) of GC and field LMXBs differ at ~4σ significance at $L_X < 5$ x $10^{37}$ erg s$^{-1}$. As previously reported, these XLFs are consistent at higher luminosities. The presently available theoretical models for LMXB formation and evolution in clusters are not sophisticated enough to provide a definite explanation for the shape of the observed GC-LMXB XLF. Our observations may indicate a potential predominance of GC-LMXBs with donors evolved beyond the main sequence, when compared to current models, but their efficient formation requires relatively high initial binary fractions in clusters. The field LMXB XLF can be fitted with either a single power-law model plus a localized excess at a luminosity of 5-6 x $10^{37}$ erg s$^{-1}$, or a broken power-law with a similar low-luminosity break. This XLF may be explained with NS-red-giant LMXBs, contributing to ~15% of total LMXBs population at ~5x$10^{37}$ erg s$^{-1}$. The difference in the GC and field XLFs is consistent with different origins and/or evolutionary paths between the two LMXB populations, although a fraction of the field sources are likely to have originated in GCs.

*Subject headings*: galaxies: elliptical and lenticular – galaxy: individual (NGC 3379, NGC 4278, NGC 4697) – X-rays: binaries – X-rays: galaxies

1. INTRODUCTION

LMXBs are luminous X-ray binaries associated with old stellar populations; they are powered by the accretion of the atmosphere of a low-mass late-type star onto a compact stellar remnant, either a neutron star or a black hole. Since their discovery in the Milky Way (see Giacconi 1974), the origin and evolution of LMXBs has been the subject of much discussion. Galactic LMXBs are found in both the stellar field and GCs, but their incidence per unit stellar mass is much higher in GCs, suggesting a dynamical formation mechanism for at least this sub-sample (Clark 1975; Katz 1975). The evolution of native binary systems is a viable, but still controversial, formation scenario for field LMXBs, which could also have been dynamically formed in GCs, and then dispersed in the field (e.g., Grindlay 1984; see reviews in Verbunt & van den Heuvel 1995, Verbunt & Lewin 2006).

*Chandra* observations have provided samples of LMXBs in many early type galaxies, rekindling the discussion of their formation and evolution. Of order 20-70% of these extra-Galactic LMXBs are found in GCs (e.g., Kundu et al. 2002; Sarazin et al. 2003; Jordan et al. 2004; Kim et al. 2006; Kundu, Maccarone & Zepf 2007, hereafter KMZ; Sivakoff et al. 2007; Humphrey & Buote 2008); as in the Milky Way, for a given stellar mass, LMXBs are more likely to be found in GCs than in the field. This result has again stimulated the hypothesis of exclusive formation in GCs for all LMXBs. However, there are also studies suggesting formation *in situ* for field LMXBs; in particular, this conclusion is supported by comparison of the LMXB population with the GC specific frequency ($S_N$) in several galaxies (e.g., Juett 2005; Irwin 2005; see Kim E., et al 2006 for cautions and Fabbiano 2006 for review and earlier references). This work was all based on the observation of the most luminous LMXBs, with $L_X$ (0.3-8 keV) ≥ a few $10^{37}$ erg s$^{-1}$. Now, deep observations of three elliptical galaxies – NGC3379 (Brassington et al. 2008a, hereafter B08a), NGC4278 (Brassington et al. 2008a, hereafter B08b) and NGC4697 (Sivakoff et al. 2008) – allow us to extend the comparison of field and GC-LMXBs to sources in the 'normal' range of Galactic LMXB luminosity.

The principal tool we use for this study is the XLF of the LMXB populations (see e.g., Kim & Fabbiano 2004, hereafter KF04; Gilfanov 2004; Fragos et al. 2008 for earlier studies of luminosity functions of X-ray sources in galaxies). The high luminosity end ($L_X$ > several x $10^{37}$ erg s$^{-1}$) of the XLF (GC+field co-added) is well constrained with a differential slope of ~1.8 (e.g., KF04; Gilfanov 2004). The normalization (i.e., the total number of LMXBs in a given galaxy) is strongly related to the stellar mass of the galaxy, although a link to $S_N$ has also been reported (White et al. 2002; Kundu et al. 2002; KF04; Kim, E. et al. 2006). KF04 and Gilfanov (2004) independently found that the XLF is broken at $L_X$ ~ 5 x $10^{38}$ erg s$^{-1}$, possibly reflecting the presence of both neutron star and black-hole LMXBs in the X-ray source populations, as suggested by Sarazin et al. (2001). This break is also predicted in the model of short-lived, high-birth-rate, ultra-compact binary evolution in GCs by Bildsten & Deloye (2004). Above the break ($L_X$ > 5 x $10^{38}$ erg s$^{-1}$), the XLF slope becomes steep (β ~ 2.8); very luminous X-ray sources (or ULX with $L_X$ > 2 x $10^{39}$ erg s$^{-1}$) are extremely rare in typical old ellipticals (Irwin et al. 2004).

The above considerations also apply to GC and field LMXB XLFs separately, since their XLFs are entirely consistent at high luminosity (Kim E. et al 2006).

In the low $L_X$ range, the (GC+field) XLF is less well characterized, because of the lack of adequately deep *Chandra* observations. Voss & Gilfanov (2006; 2007a) found that the XLFs of the LMXB populations of the nearby galaxies NGC 5128 and M31 significantly flatten below $L_X \sim 2 \times 10^{37}$ erg s$^{-1}$. However, these galaxies also contain younger sources, and some contamination of the samples cannot be excluded. Instead, using early partial observations of the 'old' elliptical galaxies NGC 3379 and NGC 4278 (110 and 140 ks, respectively), Kim D.-W. et al. (2006) found no evidence of this flattening (down to $L_X \sim 10^{37}$ erg s$^{-1}$), but suggested a possible local excess over a power law in the XLF of NGC 3379 at $L_X \sim 4 \times 10^{37}$ erg s$^{-1}$. It was also suggested, both in M31 and NGC 5128, that the XLF of GC-LMXBs may be flatter than that of field LMXBs (Voss and Gilfanov 2007a; Woodley et al. 2008). The present study seeks to establish if there is a 'universal' shape of the low-luminosity LMXB XLF in different galaxies, and if the difference suggested between field and GC XLFs is generally valid.

This paper is organized as follows. In Section 2, we describe the target galaxies, and in Section 3 the *Chandra* observations and data reduction techniques. In Section 4, we cross-correlate the X-ray and optical sources to identify GC and field LMXBs and we describe the related uncertainties in terms of contamination by foreground and background objects and chance coincidence. In Section 5, we compare the fractions of LMXBs associated with GCs and the field in each galaxy, in different luminosity ranges. We also compare field and GC luminosity distributions, including upper limits for non-detections in GCs. In Section 6, we derive the X-ray luminosity function separately for GC and field samples and we present the fitting result. In Section 7, we discuss the implications of our results for the nature of LMXBs and their formation. Finally, we summarize our conclusions in Section 8.

2. THE TARGET GALAXIES

We summarize the optical characteristics of the three target galaxies of this study in Table 1. Because these three elliptical galaxies are old (e.g., Trager et al. 2000; Terlevich & Forbes 2002), they provide a clean sample of LMXBs with no contamination by younger sources (HMXBs and SNRs). These younger sources may contaminate LMXB populations extracted from observations of spiral galaxies (the Milky Way, M31) and of young or rejuvenated E and S0 galaxies resulting from recent mergers (e.g., NGC 5128). Moreover, all three galaxies harbor only small amounts of hot ISM (see Trinchieri et al 2008, for a detailed study of NGC 3379), unlike typical X-ray bright ellipticals (e.g., M87, NGC 5128) where point sources may be confused with small-scale gas clumps, and the diffuse emission limits the detection of faint LMXBs.

We adopt distances of 10.6 Mpc (NGC 3379), 16.1 Mpc (NGC 4278) and 11.8 Mpc (NGC 4697) throughout this paper, based on the surface brightness fluctuation analysis by Tonry et al. (2001). At these distances, 1′ corresponds to 3.1 kpc, 4.7 kpc, and 3.4 kpc, respectively.

Table 1. Sample Galaxies

| galaxy (1) | D (Mpc) (2) | R25 (′) (3) | PA (deg) (4) | B_T_0 (mag) (5) | M_B (mag) (6) | L_B (L_Bo) (7) | age (Gyr) (8) | S_N (9) | N(H) ($10^{20}$ cm$^{-2}$) (10) |
|---|---|---|---|---|---|---|---|---|---|
| NGC 3379 | 10.57 | 2.69x2.39 | 67.5 | 10.18 | -19.94 | 1.46e10 | 8.6-10 | 1.2 | 2.78 |
| NGC 4278 | 16.07 | 2.04x1.90 | 27.5 | 10.97 | -20.06 | 1.63e10 | 10.7-12 | 6.9 | 1.76 |
| NGC 4697 | 11.75 | 3.62x2.34 | 67.5 | 10.07 | -20.28 | 2.00e10 | 8.2-8.9 | 2.5 | 2.14 |

1. Galaxy name
2. Distance from Tonry et al. (2001)
3. Semi-major and semi-minor axes determined at 25th magnitude from RC3
4. Position angle of the major axis from NED
5. B_T_0 from RC3
6. Absolute blue magnitude
7. Blue luminosity calculated by adopting an absolute solar blue magnitude of 5.47 mag
8. Luminosity weighted average stellar age (Trager et al. 2000; Terlevich & Forbes 2002; Thomas et al. 2005)
9. Globular cluster specific frequency from Ashman & Zepf (1998)
10. H column density along the Galactic line of sight from Dickey and Lockman (1990)

## 3. *CHANDRA* X-RAY OBSERVATIONS AND SOURCE DETECTION

NGC 3379, NGC 4278 and NGC 4697 were observed with the S3 (back-illuminated) chip of *Chandra* Advanced CCD Imaging Spectrometer (ACIS, Garmire 1997) multiple times between 2001 and 2007, with individual exposures ranging from 30 to 110 ks. NGC3379 and NGC 4278 were observed as part of a *Chandra* very large program (PI: G. Fabbiano); the archival data of NGC 4697 were obtained as part of a study by Sivakoff et al. (2008, and references therein). Observation dates and net exposure times are summarized in Table 2. In all the observations used in this study, the entire $D_{25}$ ellipse of each galaxy falls within the S3 chip, and the ACIS temperature was -120 C. We did not use an older 36 ks observation of NGC 4697 taken on Jan. 15, 2000 with detector temperature of -110 C, because of the relatively large uncertainty in calibrating the detector characteristics (http://cxc.harvard.edu/cal/Acis/).

The ACIS data were uniformly reduced in a similar manner as described in Kim & Fabbiano (2003) with a custom-made pipeline (XPIPE), specifically developed for the *Chandra* Multi-wavelength Project (ChaMP; Kim et al. 2004a). Starting with the CXC pipeline level 2 products, we apply *acis_process_event* available in CIAO v3.4 with up-to-date calibration data, e.g., time-/position-dependent gain and QE variation. We note that the proper (serial) CTI (charge transfer inefficiency) correction for the S3 (BI) chip was only applied in the CXC pipeline processing after Jan. 2007 (http://asc.harvard.edu/ caldb/downloads/Release_notes/CALDB_v3.3.0.html). After removing background flares, we re-project individual observations to a common tangent point and combine them by using *merge_all* available in the CIAO contributed package (http://cxc.harvard.edu/ ciao/ threads/ combine/). The background flares are not very significant in most observations (the exposure time reduces by less than 8%), except for the 3$^{rd}$ observation of NGC 4697 (obsid=4729), where the exposure time is reduced by 40% (or 16 ksec out of 38 ksec). The total effective exposures of the merged observations are 324 ksec, 458 ksec and 132 ksec for NGC 3379, NGC 4278, and NGC 4697, respectively. The exposure time of NGC 4697 is not as long as for the first two galaxies, but given the distances, the detection limit is comparable to that of NGC 4278. We show the merged images of the three galaxies in Figure 1, where the X-ray point sources and the optical size ($D_{25}$) are marked.

Table 2. Chandra Observations

| obsid | obs_date | exp (ksec) | | Nsrc | |
|---|---|---|---|---|---|
| | | (a) | (b) | (c) | (d) |
| **N3379** | | | | | |
| 1587 | Feb 13 2001 | 31.5 | 29.0 | 71 | 44 |
| 7073 | Jan 23 2006 | 84.1 | 80.3 | 85 | 57 |
| 7074 | Apr 9 2006 | 69.1 | 66.7 | 82 | 54 |
| 7075 | Jul 3 2006 | 83.1 | 79.6 | 85 | 57 |
| 7076 | Jan 10 2007 | 69.2 | 68.7 | 78 | 49 |
| merge | | | 324.2 | 163 | 93 |
| **N4278** | | | | | |
| 4741 | Feb 3 2005 | 37.5 | 37.3 | 96 | 58 |
| 7077 | Mar 16 2006 | 110.3 | 107.7 | 174 | 116 |
| 7078 | Jul 25 2006 | 51.4 | 48.1 | 98 | 63 |
| 7079 | Oct 24 2006 | 105.1 | 102.5 | 144 | 93 |
| 7080 | Apr 20 2007 | 55.8 | 54.8 | 120 | 74 |
| 7081 | Feb 20 2007 | 110.7 | 107.6 | 158 | 104 |
| merge | | | 458.0 | 271 | 168 |
| **N4697** | | | | | |
| 4727 | Dec 26 2003 | 39.9 | 36.6 | 75 | 68 |
| 4728 | Jan 6 2004 | 35.7 | 33.3 | 77 | 64 |
| 4729 | Feb 12 2004 | 38.1 | 22.3 | 62 | 54 |
| 4730 | Aug 18 2004 | 40.0 | 38.1 | 90 | 71 |
| merge | | | 132.0 | 129 | 102 |

Note.
a. livetime from the CXC pipeline data
b. effective exposure time after removing background flares
c. number of detected sources in the S3 chip
d. number of detected sources within the D25 ellipse

The X-ray point sources were detected using CIAO *wavdetect*. We set the significance threshold to be 10$^{-6}$, which corresponds approximately to one false source per chip and the exposure threshold to be 10% using an exposure map. The latter was applied to reduce the false detections often found at the chip edge. The performance and limitations of *wavdetect* are well understood and calibrated by extensive simulations (e.g., Kim & Fabbiano 2003; Kim et al. 2004a; Kim, M. et al. 2007a). From the merged data, we detect 163, 271 and 129 point sources in the S3 CCD chip for NGC 3379, NGC 4278 and NGC 4697, respectively (Table 2).

To measure the X-ray flux and luminosity (in 0.3-8 keV), we take into account the temporal and spatial QE variation (http://cxc.harvard.edu/cal/Acis/Cal_prods/qeDeg/) by calculating the energy conversion factor (ECF = ratio of flux to count rate) for each source in each observation. We assume a power-law spectral model with a photon index of $\Gamma=1.7$ (e.g., Irwin et al. 2003) and Galactic $N_H$ (see Table 1). To calculate the X-ray flux of sources detected in the merged data, we apply an exposure-weighted mean ECF. This will generate a flux as if the entire observations were done in one exposure, but with a variable detector QE as in the real observations. Among the five observations of NGC 3379, the ECF significantly differs only in the first observation (taken in 2001) by ~12%, while it is almost identical for the other four observations. Among the six and four observations of NGC 4278 and NGC 4697, the ECF varies only by 2% and 1%, respectively.

We note that the luminosity used in the XLF is an average value over the full observation interval. We exclude known transients (see Section 6 for the effect on the luminosity function.) We do not use X-ray sources which are detected only in one or two individual observations, but not detected in the merged data. They may be transients and their luminosities ($< 10^{37}$ erg s$^{-1}$) are below the $L_X$ range of the XLF (see Section 6).

4. SELECTION OF GC-LMXB AND FIELD LMXB SAMPLES

We used the optical source lists from Kundu & Whitmore (2001) for NGC 3379 and NGC 4278 and from Jordan et al. (2009, in preparation) for NGC 4697. Both studies utilize *HST* images to identify optical GC candidates and background galaxies. The first two galaxies were observed with WFPC2 while the latter was observed with the Wide Field Channel of ACS. We cross-correlated X-ray and optical sources to identify LMXBs in GCs and in the field, by applying strict matching criteria. We first determined the systematic positional offset between the samples of X-ray and optical sources, finding that the relative offset is <0.8″ for all three galaxies. After correcting for this offset, we assigned a match if the distance between X-ray and optical positions ($d_{XO}$) is either

(A) $d_{XO} < 0.5″$

or

(B) $0.5″ \leq d_{XO} \leq 1″$ and smaller than the X-ray positional uncertainty

While the quoted *Chandra* positional accuracy is 1″ (Chandra Proposers' Observatory Guide; http://asc.harvard.edu/proposer), *Chandra* positions are often more accurate, particularly near the aim point, as seen by comparing sources detected in multiple observations, (e.g., Chandra Deep Fields; see Kim, et al. 2004). However, the CCD pixel size (0.492″) and the mirror PSF (0.3″-0.5″ for a 50% encircled energy fraction) limit the practical minimum to be 0.5″. For faint and/or off-axis sources, the positional uncertainty of the X-ray source can be larger than 0.5″. The X-ray positional uncertainty is estimated with the empirical formula in Kim, M. et al. (2007). We take the uncertainty at a 95% confidence level. To include matches with a large positional uncertainty, we apply the 2$^{nd}$ condition (B) listed above. The matching statistics are summarized in Table 3. The chance probability of random coincidence is very low, 0.5-1.5 in each galaxy (see below).

We also consider as possible matches sources that do not satisfy the conditions A or B above, but satisfy:

(C) $d_{XO} < 2″$

We applied the above criteria to matches with either GC or BG (background galaxies) obtaining four sub-samples: (1) X-ray source (XRS) – GC matches, (2) XRS – GC possible matches, (3) XRS – BG matches, (4) XRS – BG possible matches. The number of sources in each sub-sample is listed in Table 3. We take the first sub-sample, XRS – GC matches, as GC-LMXBs. And we take only non-matches which are within the HST field of view (fov), but do not belong to any of the above four sub-samples as field LMXBs. We note that the chance probability of random coincidence among possible matches is appreciable and about half of them are real matches (see below).

Table 3. Source and Match Results

```
---------------------------------------------------------------------------------------
                              N3379    N4278    N4697     N3379    N4278    N4697
                        (all sources in the D25 ellipse) (exclude sources within r<10")
---------------------------------------------------------------------------------------
 1 XRS in D25                    93      168      102        79      154       95
 2 XRS in D25 & HST fov          59      112      102        45       98       95
 3 GC  in D25 & HST fov          70      265      449        67      257      446
 4 BGC/RGC                    30/40  144/121  195/254     29/38  140/117  194/252
 5 BG  in D25 & HST fov         346       73     1137       345       66     1134

 6 XRS-GC matches                 9       37       31         8       37       30
 7 XRS-BGC/RGC matches          4/5    12/25     7/24       3/5    12/25     7/23
 8 possible GC matches            5       14        7         3       11        7
 9 XRS-BG matches                 3        5        6         2        3        6
10 possible BG matches            6        8        9         6        5        8

11 GC fraction with LMXBs                                    12%      14%       7%
12 BGC/RGC fraction with LMXBs                            10/13%    9/21%    4%/9%

13 Field LMXBs                                                26       42       44

14 Fraction of GC-LMXBs*                                    0.24     0.47     0.40

15 Fraction of faint field LMXBs+                            81%      76%      68%
16 Fraction of faint GC-LMXBs+                               38%      43%      50%
---------------------------------------------------------------------------------------
XRS = X-ray source
GC = Globular clusters
RGC = red GC (V-I > 1.05 for NGC 3379/N4278 and g-z > 1.1 for N4697)
BGC = blue GC
BG = Background galaxies = non-GC optical sources
```

\* N(GC-LMXBs) / N(all LMXBs)
+ N(faint LMXBs with $L_X < 5 \times 10^{37}$ erg s$^{-1}$) / N(all LMXBs)

In Table 3, we list the number of sources within the $D_{25}$ ellipse in the first 3 columns. For NGC 3379 and NGC 4278, the HST WFPC2 field of view covers only a part of the $D_{25}$ ellipse (see Figure 1). For NGC 4697, the HST fov covers the entire $D_{25}$ ellipse. We do not use the sources located outside the $D_{25}$ ellipse, because they have a higher probability to be associated with foreground/background objects. The X-ray luminosity of individual sources ranges from $10^{36}$ erg s$^{-1}$ to $10^{39}$ erg s$^{-1}$ in NGC 3379 and from several x $10^{36}$ erg s$^{-1}$ to $10^{39}$ erg s$^{-1}$ in NGC 4278 and NGC 4697. The completeness also varies from one galaxy to another (see Section 6).

In the last 3 columns of Table 3, we further exclude sources inside the central region (r < 10″). In the central region, both X-ray and optical data are rendered incomplete by the strong diffuse emission and also by nearby sources particularly for the X-ray sources. Because faint X-ray sources are difficult to detect near the center, the source detection is significantly incomplete and the incompleteness is hard to measure and correct. Even if relatively bright sources are detected, their photometric quantities (and possibly their positions) are uncertain. The *HST* optical sources are also affected by similar incompleteness, because of a high background level from the host galaxy (e.g., Table 2 and 3 in Jordan et al, 2009). Moreover, both NGC 4278 and NGC 4697 are known to have central dust lanes, which make it even harder to detect GCs near the galaxy centers. Only a small number of very bright, compact GCs are found inside 10″; this result may be at least in part because of detection incompleteness. As listed in Table 3 (row 3), only 3 GCs are found inside 10″ of the center of NGC 3379 out of 70 GCs in the HST fov (8 out of 265 in NGC 4278; 3 out of 449 in NGC 4697). This is in contrast, for example, to 14 X-ray sources found inside 10″ out of 59 X-ray sources in the HST fov in NGC 3379 (14 out of 113 in NGC 4278 and 7 out of 102 in NGC 4697; row 2 in Table 3). Given that the X-ray source detection is also incomplete, the lack of GCs in the center is even more obvious. Although it is possible that a part of apparent field LMXBs might originate from the disrupted GCs, given that GCs could be disrupted more easily near the galaxy center, the incompleteness of GCs will cause more XRS identified as field LMXBs in the central region. If we had applied the same matching criteria, we would have 1 GC-LMXBs and 10 field LMXBs inside 10″ of NGC 3379 (0 vs. 6 in NGC 4278 and 1 vs. 5 in NGC 4697). We note that this is not because of the different radial profiles of GC-LMXBs and field LMXBs. We will present a full description of the radial distribution in the forthcoming paper (Kim et al., 2009 in prep.). We further divide GCs into two groups, blue and red GCs, separating them at V-I = 1.05 for NGC 3379 and N4278 and g-z = 1.1 for N4697, based on the C-M diagrams (row 4 in Table 3).

We estimated the chance coincidence of the associations by re-matching X-ray and optical sources after shifting the X-ray sources randomly. Within the *HST* fov (excluding the central 10″ region), we find the chance coincidence to be 1.5/1 for GC/BG matches in NGC 4278. The chance probabilities in the other two galaxies are lower than that of NGC 4278 which hosts the largest number of X-ray sources inside the smallest fov. Therefore, a false match in the GC-LMXB sample is extremely rare. Instead, in all three galaxies, about half of the "possible" matches may be chance associations.

About 10% of the X-ray sources are found in non-GC optical sources (or background galaxies), if we count BG matches (row 9 in table 3) and one half of the possible BG matches (row 10). The other half of the 'possible BG matches' is likely to be due to chance coincidences, resulting from the crowded source fields (see above). Based on the ChaMP+CDF log(N)-log(S) (Kim, M. et al. 2007b), we estimate the number of cosmic background sources to be 21, 12, and 17 within the $D_{25}$ ellipse of NGC 3379, NGC 4278 and NGC 4697, respectively. This is determined at the flux limit of 90% completeness (see Section 6). Cosmic background X-ray sources therefore account for 7-23% of the X-ray sources within the $D_{25}$ ellipse. The number of background sources is further reduced, if we consider only the sources found inside the *HST* fov: 5 in NGC 3379, 5 in NGC 4278 and 17 in NGC 4697. These expected numbers are almost identical to those of sources matched with BG objects, except in NGC 4697 where seven background sources possibly remain undetected. Given that the LMXBs-GCs matches are highly significant (see above), the remaining background sources will primarily contaminate the field LMXB sample by ~6% (7 out of a total 112 field LMXBs in the three galaxies).

## 5. STATISTICS OF LMXB SAMPLES

In the GC-LMXB sample (row 7 of Table 3), LMXBs are preferentially matched (by a factor 2 or more, see row 12 of Table 3), with red, metal-rich rather than blue, metal-poor GCs, in agreement with previous reports (e.g., Kundu et al. 2002; Sarazin et al. 2003; Jordan et al. 2004; Kim E. et al. 2006; Sivakoff et al., 2007). The cause of this trend is not fully understood yet, although there are a few suggested explanations (e.g., irradiation-induced stellar winds, Maccarone et al. 2004; metallicity-dependent convective zone, Ivanova 2005). Among the three galaxies, the number fraction of GC-LMXBs,

$$F_{N,\ GC\text{-LMXB}} = \frac{N_{GC\text{-LMXB}}}{N_{GC\text{-LMXB}} + N_{Field\text{-LMXB}}}$$

ranges from 25% to 50% (row 14 in Table 3). This fraction increases with increasing GC specific frequency, $S_N$ (see Table 1), as previously suggested (e.g., Juett 2005). We further discuss the $S_N$ dependency in section 6.1

At luminosities larger than $5 \times 10^{37}$ erg s$^{-1}$, the fraction of GCs associated with a LMXB is ~5% (Sarazin et al 2003; see Fabbiano 2006). This fraction increases when the detection threshold moves to lower luminosities, as first suggested by Kundu, Maccarone & Zepf (2007). An increase would be expected, extrapolating to lower luminosities the high luminosity XLF of Kim E. et al (2006); how much this fraction increases depends on the low-luminosity slope of the XLF. Comparing the fraction of GCs associated with LMXBs in two increasingly deeper exposures of NGC 3379 (that used by KMZ and the full exposure of B08a), Fabbiano (2008) noticed that this fraction does not increase, remaining at ~12-13% (for detection threshold going from $2\times10^{37}$ erg s$^{-1}$ to a few $10^{36}$ erg s$^{-1}$). Instead, the number of detected LMXBs in the field increases by a factor of 2.4, so that the fraction of LMXBs associated with GCs decreases with deeper exposures. We now find a similar effect in NGC 4278 and NGC 4697. The fraction of faint ($L_X < 5 \times 10^{37}$ erg s$^{-1}$) LMXBs is given at the bottom of Table 3 for each galaxy for both GC and field samples. Comparing GC and field faint source fractions, we find that typically there is a dearth of low-luminosity GC cluster sources, in comparison with field sources. While the faint LMXB fraction is 70-80% in the field sample (row 15 in Table 3), it is only 40-50% in the GC sample (row 16 in Table 3). Applying a proportion test, available in the R package (www.r-project.org), we find that the statistical significance of this difference in the faint LXMB fraction is at the 3.8σ level.

The dearth of low luminosity GC-LMXBs is confirmed by the results of a stacking experiment on the GCs with undetected X-ray counterparts. For this experiment, we included in the detections only sources with luminosities detected at ≥3σ confidence. We created source regions, centered on the location of the GCs, excluding those with confirmed X-ray counterparts, or too close to multiple X-ray sources for reliable photometry. Then, we performed the same aperture photometry as applied for the real X-ray sources (see KF04 and Kim et al. 2004a for photometry details). Table 4 summarizes the cumulative X-ray source and background counts (normalized to the source area) for each galaxy. Figure 2 shows the histograms of the source counts (lower panel) and background subtracted net counts (upper panel) extracted from each stacking regions; the median value of the net count distributions are 0.80, 1.91 and 0.04 for NGC 3379, NGC 4278, and NGC 4697, respectively, showing that there are no biases in the determination of the background counts.

Following the same Bayesian approach used in B08a and B08b, which takes into account the Poisson nature of the probability distribution of the source and background counts, as well as the effective area at the position of the source (Park et al 2006), we find upper bounds on the intensity of a 'stacked' source at 68% and 99.7% confidence for the three galaxies. Dividing by the number of GCs included in the three experiments, we calculate corresponding luminosity upper limits in the 0.3-8.0 keV band, with an energy conversion factor determined in the same way as for the X-ray sources (see section 3). We thus obtain upper confidence bounds on the 'average' X-ray luminosity of a non-detected GC. Since we have no way of knowing how many GCs are indeed associated with LMXBs below the detection threshold, and what their distribution of X-ray luminosities may be, we can use the results of the stacking experiment only to constrain the cumulative luminosity distribution [in cumulative $L_X( >L_{X,src})$] of GCs at the low end (not the XLFs).

Figure 3 shows the cumulative luminosity distributions of field and GC LMXBs for the three galaxies, which suggest a flattening of the cumulative luminosity distribution of GC LMXBs at low luminosities. Note that these are 'observed' distribution, not corrected for incompleteness; since both field and GC sources come from the same data and suffer from the same observational biases, direct comparison is valid. By using the stacking upper limit we can exclude that the possibility that incompleteness is responsible for the lack of GC sources at the low luminosities. Survival analysis tests (from ASURV; Lavalley, Isobe & Feigelson 1992) on these distributions show that the probabilities that GC and field populations originated from the same parent population are only 0.5% in NGC 3379 and 3-4% in the other two galaxies (Table 4).

```
              Table 4. GC Stacking Upper Limits and GC-Field comparison

      ----------------------------------------------------------------------------
                                  NGC 3379         NGC 4278         NGC 4697
      ----------------------------------------------------------------------------
      No. GCs                           56              168              433
      X-ray source counts (stacking)  1229             5864             4397
      X-ray bkg counts (stacking)    1239.3           5483.7           4109.1
      68% net counts                   28.8            463.3            355.7
      68% Lx (per GC in 10^35)          1.1             11.5              7.0
      99.7% net counts                101.1            654.6            507.4
      99.7% Lx (per GC in 10^35)       5.44            16.3             10.0
      P* (Peto-Prentice)              0.005            0.037            0.030
      ----------------------------------------------------------------------------

      * probability that two luminosity distributions of GC and field LMXBs
        come from the same parent population.
```

## 6. X-RAY LUMINOSITY FUNCTIONS OF GC AND FIELD LMXBS

To construct the XLF, we used point sources detected within the $D_{25}$ ellipse (the size and position angle are given in Table 1). Although some X-ray sources outside the $D_{25}$ ellipse may be associated with the galaxy, we excluded them to reduce the contamination by interlopers. We also excluded sources located near the galactic centers (R < 10″), because of large photometric and positional errors and difficult incompleteness corrections for both X-ray and optical data (see Section 3). With these selection criteria, we use 79, 154, and 95 sources in NGC 3379, NGC 4278 and NGC 4697, respectively. Among these sources, we identify 75 GC-LMXBs and 112 field LMXBs inside the *HST* fov (also excluding the central 10″ region) from the three galaxies, cumulatively.

To determine the XLFs accurately, it is most critical to correct for incompleteness (see Kim & Fabbiano 2003, KF04). Without this correction, the XLF would appear flattened at the lower luminosities where the detection is not complete, causing an artificial break. Following KF04, we performed extensive simulations to generate incompleteness corrections: we simulated 20,000 point sources using ***MARX*** (http://space.mit.edu/ASC/MARX/), added them one by one to the observed image and then determined whether the added source is detected. Since we used the real observed data as the baseline, we could correct simultaneously three biases: detection limit, Eddington bias (Eddington 1913) and source confusion (Kim & Fabbiano 2003). In the simulations, we assumed a typical LMXB XLF differential slope of β=2 (KF04) where β is defined in the differential form,

$$\frac{dN}{dL_X} = k L_X^{-\beta}$$

We note that the adopted XLF slope does not significantly affect the results, because the correction is determined by the ratio of the number of input sources to that of detected sources at a given $L_X$ (see also Kim and Fabbiano 2003). As shown in B08a, B08b (see also Kim E. et al. 2006; Kim et al. 2009 in prep.), the radial distribution of LMXBs closely follows that of the optical halo light, regardless of their association with GCs. Therefore, we adopted an $r^{-1/4}$ law for the radial distribution of the LMXBs. Even if the radial distribution of LMXBs deviated from that of the stellar distribution, the effect would be minimal, because we do not use LMXBs from the central regions (r<10″) where the uncertainty in the incompleteness correction obtained by using different radial profiles would be most significant.

We find that the 90% completeness limit (i.e., where 10% of sources with this luminosity would not be detected inside the $D_{25}$ ellipse, but excluding the central 10″) is $L_X = 6 \times 10^{36}$ erg s$^{-1}$ for NGC 3379, $L_X = 1.5 \times 10^{37}$ erg s$^{-1}$ for NGC 4278, and $L_X = 1.5 \times 10^{37}$ erg s$^{-1}$ for NGC 4697; we can reliably correct the XLFs to X-ray luminosities a factor of 2 lower than the 90% limit, roughly corresponding to a 30% detection limit.

To build the XLF of LMXBs in the GC and field samples separately, we combined all GC-LMXBs and field LMXBs in the three galaxies after correcting for the incompleteness in each galaxy. We discuss the XLFs of the individual galaxies in Section 6.1. We note that galaxy-to-galaxy variation is minimal, because of the similarity of our elliptical galaxies in terms of age, distance and luminosity (see Table 1). Although the *HST* coverage is different in NGC 3379/4278 and NGC 4697, i.e., we sample LMXBs from different galacto-centric radii in the three galaxies, we do not see any systematic change in the XLF as a function of radius (see below). As noted in Section 4, the chance probability of random coincidence among GC-LMXBs is very low (one or less in each galaxy) and the contamination by unidentified background objects is ~ 0% and 6% in the GC-LMXB and field LMXB samples, respectively.

To determine the XLF shape parameters, we fitted the bias-corrected XLF in a differential form with (a) a single power-law, (b) a broken power-law, and (c) a single power-law + a Gaussian function. We applied both $\chi^2$ and Cash statistics, using ***sherpa*** available in the CIAO package. The $\chi^2$ method can determine both a confidence interval of each parameter and a goodness-of-fit. To properly apply the $\chi^2$ statistic we selected the $L_X$ bin size, $\delta \log(L_X) = 0.2$, so that there is a minimum of 10 sources in each $L_X$ bin and applied the Gehrels variance function for the error calculation (Gehrels 1986). The Cash statistic (also C-stat) utilizes a maximum likelihood function and can be applied regardless of the number in each bin. In this case, we further reduced the bin size, $\delta \log(L_X) = 0.1$, to be able to identify small variations from a power-law distribution. Because in the Cash statistic the counts are sampled from the Poisson distribution in each bin, we could not apply the correction to the observed XLF before the fit. Instead, we fitted the uncorrected XLF with the modified model, which is divided by the correction factor. When we plot the XLF, the correction factor is multiplied back to the model. Both statistics result in consistent parameters within the error. We present the fitting results from both statistics in Table 5. We take the best-fit parameters from the Cash statistic and the goodness-of-fit from the $\chi^2$ statistic.

We show the combined, bias-corrected XLFs for the field and GC samples in Figure 4 (the best fit model is from the Cash statistic). The differential XLF is plotted in the form of $dN/d\ln L_X$ as a function of $L_X$ (instead of $dN/dL_X$ vs. $L_X$). In this form, the slope, if a single power-law is applied, will be the same as that of the cumulative XLF so that the XLF is easily visualized and compared (e.g., Voss & Gilfanov 2007a). As is clearly seen in Figure 4, the XLFs of the field and GC samples differ: the GC XLF has a considerably flatter slope than the field XLF. The significance of the difference in the XLF slope is ~5σ, when a single power-law is used. The difference is more significant at lower luminosities ($L_X < 5 \times 10^{37}$) than at higher luminosities. If a broken power-law is used, the XLF slopes are consistent within ~2σ at high luminosities ($L_X > 5 \times 10^{37}$) with a slope of β ~ 2. But at low luminosities ($L_X < 5 \times 10^{37}$), the significance of the difference is ~3.5σ. While the XLF of field LMXBs continues to go up to the lowest $L_X$, the XLF of GC-LMXBs flattens to $dN/d\ln L_X$ ~ constant (or β ~ 1). This is fully consistent with finding a considerably lower fraction of faint LMXBs in the GC sample, as discussed in Section 5.

A more careful look at the XLFs suggests another interesting feature: there may be an excess over a single power law at $L_X = 5$-$6 \times 10^{37}$ erg s$^{-1}$. While this 'bump' had been suggested by the XLF of NGC 3379 (Kim D.-W. et al 2006), this feature was not statistically significant due to the limited number of LMXBs. It is seen more clearly in the field LMXB sample than the GC-LMXB sample (Figure 4). If we fit with a single-power + a Gaussian component, the Gaussian component corresponds to 17% of total field LMXBs in number. In this case, the best fit slope is ~1.8 in $L_X$ ranging from $8 \times 10^{36}$ erg s$^{-1}$ to several $\times 10^{38}$ erg s$^{-1}$. We note that this slope is the same as the best-fit slope determined with a larger sample (but limited to $L_X >$ several $\times 10^{37}$ erg s$^{-1}$) by KF04.

For comparison with previously published (GC plus field) LMXB XLFs, we made a combined XLF with all the LMXBs detected within $D_{25}$ (but at r > 10″) of the three galaxies, although we are aware that we are mixing two different types of XLFs. Again we fit the combined XLF with three different models (see Figure 5 and Table 5). The single power-law model is now clearly rejected ($\chi^2_{red}$ = 1.6 for 9 dof) and an excess at ~6 $\times 10^{37}$ is clearly seen. The other two models fit the data equally well. The single-power law + Gaussian model is slightly better than the broken power-law model, but we cannot statistically distinguish between them. In the broken power-law model, the low-luminosity break is at $L_X = 6$-$8 \times 10^{37}$ erg s$^{-1}$ and the slopes are 1.4 and 2 below and above the break, respectively. This result follows the general trend seen in M31 and NGC 5128 by Voss and Gilfanov (2006, 2007a). However, there is a quantitative disagreement in that our break luminosity is higher than the $2 \times 10^{37}$ erg s$^{-1}$ determined by Voss & Gilfanov (2006; 2007a) and our XLF slope below the break is also steeper than the slope ~1 determined by Voss & Gilfanov. Alternatively, the XLF may be represented by a single power-law for $L_X$ ranging from $8 \times 10^{36}$ erg s$^{-1}$ to several $\times 10^{38}$ erg s$^{-1}$, with a localized Gaussian 'bump', marked by the blue dashed histogram in Figure 5. The Gaussian peaks at $L_X = 5 \times 10^{37}$ erg s$^{-1}$ and has a FWHM of $7 \times 10^{37}$ erg s$^{-1}$. This component corresponds to 15% of total LMXBs (the blue histogram in Figure 5).

Also seen is a deficit at $L_X > 5 \times 10^{38}$ erg s$^{-1}$, consistent with the previously reported higher-luminosity break (KF04; Gilfanov 2004). While we have 5 sources with $L_X > 5 \times 10^{38}$ erg s$^{-1}$ (as used in the fit; Figure 5), if the XLF continued with the same slope to higher luminosity, we would expect 25 and 15 sources with $L_X > 5 \times 10^{38}$ erg s$^{-1}$, for a single power-law + bump and a broken power model, respectively. We do not plot this extra break in Figure 5, because it is not necessary in a differential XLF form. However, one should take this high luminosity break into account, if a cumulative XLF is plotted.

A few transient candidates are identified in each galaxy (see B08a and B08b for NGC 3379 and NGC 4278, respectively; we applied the same technique for NGC 4697). One of five transient candidates in N3379 is identified as a field LMXB, another one as a possible BG and the remaining three sources are out of the HST fov; one of three transient candidates in NGC 4278 is identified as a GC-LMXB, and two are in the field; one transient candidate is identified in NGC 4697 and may be a field LMXB, but it is outside of the $D_{25}$ ellipse. We have re-built the XLF without these transients, but the results do not change in any significant manner. As described in Section 4, about 6% of the field LMXB sample may be contaminated by background sources (after excluding known background galaxies). Adding this background component to the model, the results do not change, because the contamination is very small and because the logN-logS relationship of the cosmic X-ray background sources (with β = 1.7 below the break which corresponds to $L_X = 3 - 7 \times 10^{38}$ erg s$^{-1}$ at the distance of three galaxies; Kim, M. et al. 2007b) is almost identical with the XLF shape of field LMXBs.

Table 5. XLF Parameters

```
                   Field LMXBs
-------------------------------------------------------
                chi2               cash
-------------------------------------------------------
(a) a single power-law
beta         1.72 (-0.08, +0.07)  1.70 (-0.06, +0.06)
Chi2_red     1.05 ( 8.4 / 8)

(b) a broken power-law
beta1        1.23 (-0.09, +0.07)  1.27 (-0.06, +0.06)
beta2        2.95 (-0.42, +0.78)  2.48 (-0.24, +0.27)
Lx(break)*   0.60 (-0.10, +0.11)  0.55 (-0.07, +0.08)
Chi2_red     0.34 ( 2.1 / 6)

(c) a single power-law + Gaussian
beta         1.80 (-0.08, +0.08)  1.82 (-0.06, +0.06)
gauss pos*   0.60 (-0.25, +0.03)  0.60 (-0.22, +0.03)
gauss fwhm*  0.42 (     , +0.19)  0.39 (-0.13, +0.13)
Chi2_red     0.38 ( 1.9 / 5)
-------------------------------------------------------
```

```
                     GC-LMXBs
-------------------------------------------------------
                 chi2                cash
-------------------------------------------------------
(a) a single power-law
beta       1.25 (-0.13, +0.11)  1.23 (-0.09, +0.09)
Chi2_red   0.23 ( 1.9 / 8)

(b) a broken power-law
beta1      0.98 (-0.08, +0.07)  0.88 (-0.11, +0.10)
beta2      1.96 (-0.27, +0.40)  1.61 (-0.16, +0.18)
Lx(break)* 1.0  (-0.23, +0.53)  0.68 ( 0.13, +0.18)
Chi2_red   0.25 ( 1.5 / 6)

(c) a single power-law + Gaussian
beta       1.25 (-0.14, +0.12)  1.24 (-0.10, +0.09)
gauss pos* 0.55 ( 0.00, +0.00)  0.62 (-0.02, +0.02)
gauss fwhm* 0.4                  0.4
Chi2_red   0.22 ( 1.3 / 6)
-------------------------------------------------------
(gauss fwhm is fixed)

                     All LMXBs
-------------------------------------------------------
                 chi2                cash
-------------------------------------------------------
(a) a single power-law
beta       1.64 (-0.04, +0.04)  1.55 (-0.04, +0.04)
Chi2_red   1.60 (14.4 / 9)

(b) a broken power-law
beta1      1.35 (-0.05, +0.05)  1.31 (-0.04, +0.04)
beta2      2.26 (-0.20, +0.27)  2.01 (-0.12, +0.13)
Lx(break)* 0.86 ( 0.16, +0.16)  0.68 (-0.09, +0.11)
Chi2_red   0.66 ( 4.6 / 7)

(c) a single power-law + Gaussian
beta       1.72 (-0.05, +0.05)  1.61 (-0.04, +0.04)
gauss pos* 0.44 (-0.12, +0.11)  0.57 (-0.06, +0.06)
gauss fwhm* 0.80 (-0.20, +0.21) 0.41 (-0.11, +0.13)
Chi2_red   0.47 ( 2.8 / 6)
-------------------------------------------------------

* Lx(break) and gauss pos and fwhm (full width half max)
  are in units of 10^38 erg s^-1.
```

We tested whether the XLF varies as a function of galacto-centric distance, as suggested in the inner ($r < 1'$ or $r < 200$ pc) bulge of M31 by Voss & Gilfanov (2007b). However, we note that our test applies to a larger scale (in order of a few kpc) than that in M31. We divided sources at $r_b = 45$-$60''$ to make two sub-samples ($10'' - r_b$ and $r_b - D_{25}$) with a similar number and separately applied the bias-correction to each sub-sample, because the incompleteness is different in different regions. We find no statistically significant difference in best fit parameters as they are consistent with each other within the statistical errors. We repeated the same test by dividing two samples at the effective radius ($r_b = r_e$), but the results do not change. We also compared the number of luminous sources in the inner and outer regions (separated by $r_b$). Although slightly more luminous sources ($L_X > 5$ x $10^{37}$ erg s$^{-1}$) are found in the inner region, the difference is not significant within the errors.

6.1 The XLFs of the individual galaxies and $S_N$ dependencies

We show the XLFs from each galaxy in Figure 6, The XLFs do not vary much from one galaxy to another. In every case, the XLF parameters (except for the normalization, see below) are consistent within the error with each other and with the combined XLF. Faint LMXBs are preferentially found in the field sample, compared to bright LMXBs, as was seen in the combined XLF. The excess number of LMXBs is also seen in all three individual galaxies (in the left panel) in the luminosity range of $L_X = 5 - 8$ x $10^{37}$ erg s$^{-1}$, as shown in Figure 5c, but the significance is less than in Figure 5c. This may be seen in the field sample (middle panel), again with less significance than in Figure 4a.

We note that if the sample (without distinguishing between GC and field LMXBs) were only complete down to $L_X = 5$ x $10^{37}$ erg s$^{-1}$ (i.e., the bright half of XLF in the left panel of Figure 6), the XLF would be best described by a single power-law with a slope close to β=2, as determined with previous shallow data (e.g., KF04; Gilfanov 2004).

Table 6 GC-LMXB vs. Field LMXB*

```
------------------------------------------------------------------------------------------
                   |           N(LMXB)                   |           Lx(LMXB)**
       Lₖ    Sɴ    |    measured          expected       |    measured          expected
                   | GC : Field (ratio) all  GC : Field  | GC : Field (ratio) all  GC : Field
      (1) (2a)(2b)|  (3)   (4)    (5)   (6) (7)   (8)    |  (3)   (4)   (5)   (6) (7)   (8)
------------------------------------------------------------------------------------------
N3379 7.4 1.2 0.5 |  6  :  13   (0.46)   37  12 :  25    | 0.10 : 0.06  (1.7)  0.28  0.18 : 0.11
N4697 8.4 2.5 1.5 | 24  :  39   (0.62)   77  29 :  48    | 0.31 : 0.20  (1.6)  0.74  0.45 : 0.29
N4278 7.4 6.9 3.6 | 33  :  31   (1.06)  112  58 :  54    | 0.32 : 0.23  (1.4)  0.93  0.54 : 0.39
------------------------------------------------------------------------------------------
```

\* Only LMXBs with $L_X > 1.5 \times 10^{37}$ erg s$^{-1}$ and 10" ≤ r ≤ $D_{25}$ are considered here.
\*\* $L_X$(LMXB) in units of $10^{40}$ erg s$^{-1}$.
1. $L_K$ in units of $10^{10}$ $L_{\odot K}$ (assuming the absolute K mag of the Sun = 3.33 mag)
2a. GC specific frequency (repeated from Table 1)
2b. GC specific frequency determined locally within HST fov (see text)
3. GC-LMXBs identified within the HST fov and in 10" ≤ r ≤ $D_{25}$
4. field-LMXBs identified within the HST fov and in 10" ≤ r ≤ $D_{25}$
5. Ratio of GC-LMXBs to field-LMXBs
6. all LMXBs in 10" ≤ r ≤ $D_{25}$ (but excluding those matched with non-GC optical sources)
7. expected GC-LMXBs in 10" ≤ r ≤ $D_{25}$, assuming the same ratio in column 5
8. expected field-LMXBs in 10" ≤ r ≤ $D_{25}$, assuming the same ratio in column 5

The most significant difference in XLF from one galaxy to another is in the normalization. While the three galaxies are similar in their stellar luminosities, $L_K = 7.4 - 8.4$ x $10^{10}$ $L_{K\odot}$ (see Table 6), the total X-ray luminosity of LMXBs, $L_X$ (LMXB) varies by a factor 2, from 7 x $10^{39}$ and 1.3 x $10^{40}$ erg s$^{-1}$, by counting all point sources detected within the $D_{25}$ ellipse. Here we include those inside 10″ (but exclude the nuclear source at the galaxy center) and incompleteness is not corrected. The corresponding ratio of X-ray to K-band luminosities is $L_X$ (LMXB) / $L_K$ = 1 – 2 x $10^{29}$ erg s$^{-1}$ / $L_{K\odot}$, similar to those in KF04. As discussed in KF04, the X-ray to K-band luminosity ratio increases with increasing GC specific frequency ($S_N$), i.e., the luminosity ratio is largest in NGC 4278 and smallest in NGC 3379.

To further quantify the trend, we measure the ratio of GC and field LMXBs in number and X-ray luminosity from each type of LMXBs, by assuming the same ratio of GC and field-LMXBs identified within the *HST* fov to extrapolate to the entire galaxy within the $D_{25}$ ellipse (but excluding the central 10″ region). We consider only LMXBs with $L_X$ > 1.5 x $10^{37}$ erg s$^{-1}$ for homogeneous completeness and we exclude those matched with non-GC optical sources (background galaxies), again assuming the same BG fraction inside and outside of the *HST* fov. The number ratio of GC to field LMXBs ranges from 0.5 to 1, increasing with increasing $S_N$. The luminosity ratio is higher (~1.5), since most GC-LMXBs are brighter than field-LMXBs. However, we note that the luminosity ratio does not vary from one galaxy to another. This may be partly because a few very bright sources dominate. For example, the ULX in NGC 3379 could significantly change the total luminosity (it is excluded since it is within 10″).

Since the GC specific frequency may increases with increasing galacto-centric distance, we use the local $S_N$ which was determined within the HST fov (see column 2b in Table 6). We take the local $S_N$ for NGC 3379 (0.5) and NGC 4278 (3.6) from Kundu & Whitmore (2001) and estimate the local $S_N$ of NGC 4697 to be 1.5, based on the total detected GCs within $D_{25}$ and $M_V$=-21.22 (*V* from RC3). We note that although the WFPC2 fov does not cover the entire $D_{25}$ ellipse, its coverage is almost like a pie (see Fig 1) so that the effect of sampling GCs in different distances would be minimal. When compared to global $S_N$ (see column 2a in Table 6), the local $S_N$ is about a factor 2 lower uniformly for all three galaxies.

The trend that the number and luminosity of LMXBs increases with increasing $S_N$ is more significant in the GC-LMXB sample than in the field LMXB sample (see Figure 7). For example, $N$(LMXB)/$L_K$ and $L_X$ (LMXB)/$L_K$ for GC-LMXBs vary by a factor 3-5 between NGC 3379 and NGC 4278. It is suggestive that $N$(GC-LMXB)/$L_K$ may vary linearly with $S_N$. In this case, the best fit relation, as shown by a dashed line in Figure 7a, is

$$N(\text{GC-LMXB}) / L_K \, (10^{10} \, L_{K\odot})^{-1} = 2.2 \, (\pm 0.9) \, S_{N,local}$$

This relationship needs to be confirmed with a large sample of galaxies. It is also interesting to note that in the field-LMXB sample, $N$(LMXB)/$L_K$ and $L_X$ (LMXB)/$L_K$ increase (by a factor of two) with increasing $S_N$, although it is a weaker dependence than observed for GC-LMXBs. This is in contrary to what is expected if field LMXBs are totally independent from GCs. We discuss the implications of this trend in Section 7.3.

7. DISCUSSION

Our study of the field and GC-LMXB populations of three old elliptical galaxies (NGC 3379, NGC 4278 and NGC 4697) with deep Chandra observations, is in agreement (Section 5) with the previously reported preferential association of LMXBs with red GCs (e.g., Sarazin et al. 2003; Jordan et al. 2004; Kim E. et al. 2006; Sivakoff et al. 2007; KMZ), and confirms (Section 6.1) the relation of the number of GC-LMXB associations with GC specific frequency (White et al. 2002; KF04; Juett 2005). Moreover, comparing the luminosity distributions of GC and field LMXBs in the three galaxies and their XLFs (Sections 5 and 6), we find: (1) a relative dearth of GC-LMXB associations at 0.3-8~keV luminosities lower than ~5-6×$10^{37}$ erg s$^{-1}$, and (2) a break at a similar luminosity, or possibly a localized source excess, in the XLF of field LMXBs.

In the following we will discuss our results and their implication for our understanding of LMXB evolution.

7.1 The relative lack of low-luminosity GC-LMXBs – Constraints on GC binaries

The XLF of the GC-LMXBs is flatter (3.5σ significance) than that of the field LMXBs for $L_X$ < 5-6 x $10^{37}$ erg s$^{-1}$ (Figure 4). Instead, for $L_X$ > 5-6 x $10^{37}$ erg s$^{-1}$, the GC and field XLFs are consistent within 2σ, as previously reported (e.g., Kim E. et al. 2006). In this higher luminosity range ($L_X$ > 5 x $10^{37}$ erg s$^{-1}$), we find that the fraction of GC associated with LMXBs is ~5% (41 out of 769), similar to that observed in other elliptical and S0 galaxies for a comparable luminosity threshold (KMZ, see also Fabbiano 2006 and references therein). The flattening of the GC-LMXB XLF at low luminosity is consistent with earlier suggestions in the study of LMXB populations in NGC3115 (KMZ), in M31 and the Milky Way (Voss & Gilfanov 2007a), in Virgo early type galaxies (Sivakoff et al. 2008), and in NGC 5128 (Woodley et al. 2008). Our results, based on homogeneous old stellar population elliptical galaxies, which do not suffer from the distance uncertainties of Galactic sources, and from possible contamination of younger binaries (and background galaxies) as in M31 and NGC 5128, suggest that this behavior may be a general feature of LMXB populations.

Why do the GC and field XLFs differ? Given the uncertainties in the GC-LMXB XLF (Figure 4), the observations may be explained with either an excess of luminous sources or with a lack of low-luminosity sources. We will address both possibilities in turn.

A relative excess of high luminosity LMXBs in GCs may occur because of the expected overwhelming presence of transients at the high luminosities in the old stellar field population of elliptical galaxies (Piro & Bildsten 2002; King 2002). On the other hand, GC binaries with either main sequence (MS), red giant (RG), or white dwarf (WD) donors can be bright, persistent X-ray sources, because they form predominantly by stellar interactions (Clark 1975; Katz 1975; Ivanova et al. 2008) and so escape the age constraints of primordial field binaries. While there are processes leading to high-luminosity transients in GCs, they are expected to be rare. These include the capture of a RG star (rare because of their short lifetimes), or a BH+MS binary evolving to BH+RG (rare because the nuclear evolution has to overcome angular momentum losses). Short-period transients in the field have nuclear-evolved companions because of complex previous evolution stages that do not apply to GC sources made by dynamical stellar interactions. Even black hole (BH) sources in GCs could be persistent (Kalogera, King & Rasio 2004). High luminosity GC-LMXBs may be BH binaries, given their luminosities near or above the Eddington limit of an accreting NS and their widespread variability (this paper and Sivakoff et al. 2007; Maccarone et al 2007). However, some high-luminosity transients exist in GCs: B08b report a high luminosity GC transient candidate in NGC 4278, and five of the 13 luminous GC X-ray sources in the Galaxy are transients (see Verbunt & Lewin 2006, although these Galactic sources are not as luminous as those discussed here).

A relative lack of low-luminosity GC-LMXBs may be the result of observational bias, or a real effect. The former, which we can discount on account of source variability (see also Sivakoff et al 2007), could be due to multiple (confused) LMXBs in the most luminous GCs that might 'remove' sources from the fainter portion of the XLF. The majority of GC-LMXBs with $L_X \geq 10^{38}$ erg s$^{-1}$ are variable. In NGC 3379, two of the three most luminous GC-LMXBs, with $L_X \geq 1\times10^{38}$ erg s$^{-1}$, are highly variable between observations (B08a); in NGC 4278, six out of the 10 sources with $L_X \geq 1\times10^{38}$ erg s$^{-1}$ are variable, with variability up to a factor of 4 (B08b); using the same criteria for NGC 4697, we find that of the 11 sources with $L_X \geq 1\times10^{38}$ erg s$^{-1}$, eight are variable, with variability up to a factor of ~3 (see also Sivakoff et al 2008). The latter, more likely effect may result from the transition from persistent to transient X-ray sources at low luminosities. This conclusion is supported by the low limits on the luminosity of undetected GC sources found in our stacking experiment (Section 5). In the disk instability model (King et al 1997) this transition occurs when the mass transfer rate driven from the donor drops below a critical value. Since in persistent X-ray sources the mass transfer rate is thought to be directly connected to the X-ray luminosity, this transition would lead to a dearth of X-ray sources with luminosity lower than the one corresponding to the critical mass transfer rate. Although the existence of low-luminosity transients is hard to establish, because of statistical constraints (see B08a, b), we have instances of highly variable low-luminosity sources, which disappear in some observations. In NGC 3379 we detect a possible transient candidate with $L_X < 1 \times 10^{38}$ erg s$^{-1}$ (B08a).

Ultimately to understand the physical reason for the lack of low-luminosity sources in the clusters, we need reliable theoretical models for the formation and evolution of X-ray binaries in globular clusters, that account for all dynamical interactions, as well as binary evolution processes. In the current absence of such models, in what follows we discuss a number of possible explanations in view of the limited theoretical models available in the literature at present. X-ray binary populations are thought to consist of three main sub-populations based on the donor type: degenerate white dwarf, main sequence and giant stars.

Bildsten & Deloye (2004) first suggested that GC-LMXBs may be dominated by ultra-compact binaries (UCs) with NSs accreting from WD companions. They showed that persistent UCs must have a high-luminosity XLF consistent with the observed high luminosity XLFs of both GC and field LMXBs (see KF04; Kim, E. et al 2006). However, given the thermal disk-instability model (e.g., Frank, King, & Raine 2002), quantitative consideration of the cut-off X-ray luminosity for UCs with He-rich donors leads to cut-off values lower than those indicated by our measurements. Figure 11 in Deloye & Bildsten (2003) indicates a value of ~ $5\times10^{36}$ erg s$^{-1}$ for non-irradiated accretion disks, a factor of 10 lower than our observed cut-off, and the cut-off would occur at significantly lower values for irradiated disks (see, for example, King, Kolb, & Burderi 1996). However, there is overwhelming evidence that Galactic LMXB accretion discs are significantly irradiated: van Paradijs & McClintock (1994) showed that irradiation fixes their absolute visual magnitudes, and van Paradijs (1996) and King, Kolb & Burderi (1996) that irradiation determines whether these systems are transient or not. There is of course no reason why this conclusion should change for the GC-LMXBs. Accordingly, we tentatively conclude that the observed cut-off luminosity identified here is not consistent with the suggestion that the majority of the GC-LMXBs are UCs. We would be amiss however, not to mention that the sample of Galactic UCs may indicate that the cut-off X-ray luminosity might be as high as ~5 x $10^{37}$ erg s$^{-1}$ (N. Ivanova 2009, private communication).

Models of the thermal disk instability for irradiated H-rich material, disk sizes corresponding to orbital periods in the range 10-24hr change from persistent to transient behavior at $L_X$ ~ 1-3 x $10^{37}$ erg s$^{-1}$ (Dubus et al. 1999). With moderate irradiation these values could shift closer to the observed break of 5 x $10^{37}$ erg s$^{-1}$. Current evolutionary models though show that LMXBs with MS donors have typical orbital periods of less than 10hr (Fragos et al 2008), and hence cannot account for the observed cut-off at 5 x $10^{37}$ erg s$^{-1}$. Such high luminosities could be possible with strong magnetic braking (as adopted, e.g., by Stehle, Kolb & Ritter 1997 for Pop II Galactic systems). However, significantly weaker magnetic braking prescriptions are favored currently (see Ivanova & Taam 2003 and Fragos et al 2008), leading us to conclude that LMXBs with MS donors could not account for the XLFs observed above ~$10^{37}$ erg s$^{-1}$.

The last possibility is that the cluster population is dominated by persistent LMXBs with RG donors, with a truncated orbital period distribution due to cluster interactions. We note that, for such systems at a typical orbital period of about 1 day (Fragos et al 2008), the transition from persistent to transient behavior is expected to occur at 3 x $10^{37}$ erg s$^{-1}$ (Dubus et al. 1999), which also comes close to the observed break. It important to note however that LMXBs with RG donors are typically formed though binary-single and binary-binary encounters (Ivanova et al 2008); for this channel to have a significant LMXB formation efficiency, significant initial binary fractions in clusters are needed, but appear to be disfavored by the most recent dense cluster observations (Davies et al 2008). Last, we should also note that the high-end XLF may have contributions or even be dominated by LMXBs with BHs, but at present no theoretical cluster models have accounted for this possibility. Ultimately, deciding which of these populations dominates cluster LMXBs would require self-consistent modelling of LMXB formation and evolution in clusters.

7.2 Field LMXB XLF – is there a RG-LMXB signature?

As discussed in Section 6, the field LMXB XLF can be fitted with either a power-law broken at ~5 x $10^{37}$ erg s$^{-1}$, or with a single power-law (also with a high luminosity break) and a localized bump, modeled with a Gaussian component peaking at $L_X$ = 5-6 x $10^{37}$ erg s$^{-1}$ with a FWHM of 7 x $10^{37}$ erg s$^{-1}$ and accounting for ~15% of the total number of LMXBs.

The synthetic XLF models of Fragos et al. (2008) as discussed above suggest that the feature at $L_X$ = 5-6 x $10^{37}$ erg s$^{-1}$ may be caused by a single type of LMXB, persistent NS - RG binaries (red solid line Fig 5d, converted from Fig 2a in Fragos et al. 2008). As a homogeneously old stellar system, our elliptical galaxies would host RGs of a uniform age, ~10 Gyr., corresponding to a narrow mass range, ~1 M$_\odot$, which would produce a narrow range of X-ray luminosity in a binary when they start mass transfer above the critical rate to be persistent (see equation 5 in Fragos et al. 2008). Transient RG LMXBs (red, dashed line in Fig 5d) may also contribute in this luminosity range (Fragos et al. 2009). Only a small number of LMXBs are identified as transients (B08a, B08b, and present paper), but the sensitivity of the observations is relatively poor (even for such long observing times, given the small collecting area of Chandra); moreover, the time monitoring of the galaxies is limited to a few visits in a few year period and on-states may be long, therefore the effects of persistent and transient sources at $L_X$ = 5-6 x $10^{37}$ erg s$^{-1}$ cannot be separated observationally. Theoretical models of field LMXB evolution, however, stress the importance of the transient population and point out that they can have a dominant effect, depending on what determines the transient duty cycles (see Piro and Bildsten 2002; Fragos et al. 2009).

Last, we investigated whether the $L_X$ = 5-6 x $10^{37}$ erg s$^{-1}$ feature may be due to obscured LMXBs whose $L_X$ is close to the Eddington luminosity, detected at lower luminosity because of absorption from Eddington-induced outflow. We dismiss this hypothesis, because we do not find significant differences in either hardness ratios or X-ray colors between LMXBs in the 'bump' and the rest of the sample (see B08a, b for hardness ratios and colors for detected sources).

7.3 Formation of LMXBs: Field and Clusters

We will discuss here how our results bear on the long-standing controversy of LMXB formation and evolution (e.g., Grindlay 1984; Grindlay & Hertz 1985; review by Verbunt & van den Heuvel 1995): dual formation paths in GC and the stellar field, or a single path in GC, with subsequent dispersal in the field. The detection of GC sources with *Chandra* (see review, Fabbiano 2006) and the correlation of the ratio of the integrated LMXB X-ray luminosity over the total stellar luminosity of the galaxies with GC specific frequency $S_N$ (KF04) clearly shows that GC formation is important; however, the overall correlation between integrated LMXB luminosity and stellar luminosity (Gilfanov 2004; KF04) also suggest a link of the number of LMXBs to the mass of the galaxy, and therefore to the evolution of long-lived field binaries. More recently, the dual evolution hypothesis has gained support from the observations of the Sculptor dwarf spheroidal galaxy (Maccarone et al 2005), suggesting that the binary properties of field and GC-LMXBs might be different (see also KMZ); and from work based on LMXB and GC population statistics in elliptical galaxies (Irwin 2005; Juett 2005). Detailed comparisons of the radial profiles of GC and field LMXBs have proven inconclusive, because both samples follow the radial distribution of the stellar light, at least excluding the centermost, possibly confused, regions (Kim et al 2006; Humphrey & Buote 2008; KMZ). Similarly, at high luminosity the XLFs are consistent (Kim, E. et al 2006).

The present work clearly shows a difference between field and GC-LMXB XLFs at $L_X < 5\times10^{37}$ erg s$^{-1}$. This result is certainly consistent with a dual evolution path for the GC and field LMXB populations. We also note that the observed number of field LMXB can easily be produced with the evolution of native binaries in the field (see the population synthesis of Fragos et al 2008), and that from a theoretical standpoint the GC XLF may be explained with dynamically formed binaries (see Section 7.1).

We have shown in Section 6.1 that the number of GC-LMXB in our three galaxies is a strong function of the GC specific frequency $S_N$. If the field sample all originates from the evolution of primordial binaries, we would not expect any dependence on $S_N$ of the number of field LMXB; however, we find such a dependency, albeit weaker than for the GC sample. This result suggests that a fraction of the field LMXBs might have been formed dynamically in GCs, and are now found in the field because they were ejected from GCs or the parent GCs were disrupted. In NGC 4278, as much as half of the current field LMXBs may have formed dynamically in GCs (Table 6). Future comparison with larger samples of galaxies observed with both *HST* and *Chandra* are needed to put this result on a stronger statistical footing.

8. SUMMARY AND CONCLUSIONS

This paper reports the result of our study of the LMXB populations of three nearby elliptical galaxies with deep *Chandra* ACIS observations and with good optical *HST* coverage: NGC 3379, NGC 4278, and NGC 4697. Using the field of view covered by both observatories we have identified 75 GC and 112 field LMXBs within the $D_{25}$ ellipses of the galaxies and excluding the crowded inner 10″ from the galaxy centers. The co-added populations are 90% complete down to luminosities in the range of $6 \times 10^{36} - 1.5 \times 10^{37}$ erg s$^{-1}$.

With these data:

1. We confirm (Section 5) previous reports of preferential association of LMXBs with red rather than blue GCs (e.g., Kundu et al 2002, see Fabbiano 2006 and references therein).

2. Of order 32-52% of the LMXBs are associated with GCs (Table 6), and the fraction is larger in galaxies with a larger GC specific frequency $S_N$, consistent with previous reports (e.g. KF04; Juett 2005; Irwin 2005). However, because of our optical identifications we can extend this comparison to GC-LMXB and field-LMXB separately. While there is a stronger $S_N$ dependence on the number fraction in the former, using XLFs of individual galaxies (Section 6.1), we still observe a weaker-dependence in the latter, suggesting that a fraction at least of field LMXBs may have originated in GCs.

3. The relative amount of GC to field LMXBs decreases (with 3.8σ significance) at low luminosities $L_X \leq 5\times10^{37}$ erg s$^{-1}$. This result is reinforced by a stacking experiment that sets constraints on the average X-ray luminosity of undetected GCs (Section 5).

4. The co-added GC and field LMXB XLFs (Section 6) differ, at low luminosity, in agreement with the above conclusion, with the GC-LMXB showing a remarkable flattening below $L_X \leq 5\times10^{37}$ erg s$^{-1}$. The field XLF also shows a feature at the same luminosity, which can be alternatively be fitted with a less severe break or with a localized excess over a single power law distribution. The overall (GC plus field) XLF shows this low luminosity break (or feature) and is consistent at high luminosity with previous reports (Gilfanov 2004; KF04) of a break at ~$5\times10^{38}$ erg s$^{-1}$.

5. Variability is widespread in the high-luminosity GC sample, suggesting that these detections are likely to be dominated by single luminous sources, instead of multiple less luminous sources.

These results have the following implications (Section 7):

1. The difference between GC and field XLF could be due to a difference in the high-luminosity population, if these sources are persistent in GC and transient in the field. However, reports of transients in high luminosity GC sources weaken this possibility. Alternatively, there must be a genuine lack of low-luminosity sources in GCs, which can be explained with low-luminosity transients. The break luminosity is inconsistent with theoretical prediction for the critical luminosity of UC binaries (Bildsten & Deloye 2004); however, empirical evidence from the Milky Way may indicate transient UC at the observed break luminosity of ~5 x $10^{37}$ erg s$^{-1}$ (N. Ivanova 2009, private communication).

2. Persistent GC-LMXBs with MS companions are likely to have $L_X < 10^{37}$ erg s$^{-1}$. (Fragos et al 2008), and therefore are unlikely to contribute to the cluster XLF, unless mass transfer in these bright systems is driven by magnetic braking, in the standard form for Pop II systems (Stehle, Kolb & Ritter 1997). LMXBs with H-rich giant donors and moderately irradiated disks change from persistent to transients at an X-ray luminosity consistent with the observed break. However, their efficient formation requires significant initial binary fractions, that may not exist in dense globular clusters.

3. Comparison of field LMXB XLF with field evolution population synthesis models shows that the break of the XLF at ~5 x $10^{37}$ erg s$^{-1}$ may be explained by the contribution of RG sources (Fragos et al 2008). Although these RG binaries are expected in the field, they may only contribute marginally to the GC LMXB population.

4. Overall, our results are consistent with a dual formation channel for LMXBs: dynamical formation in GCs and evolution of native field binaries, although some of the binaries detected in the field may have had a dynamical origin in GCs as well, as suggested by the $S_N$ dependence.

Concluding, our results demonstrate the power of sensitive high-resolution observations of galaxies for investigating the evolution of their X-ray source populations. In the future we propose to pursue these investigations, by extending the correlations of GC and field LMXB sample with $S_N$ to a larger sample of galaxies observed with Chandra and Hubble. Our data provide several robust observational constraints for theoretical simulations of dynamical and primordial binary evolution. It is clear that more sophisticated theoretical models are necessary to provide the answers to the questions posed by the current observations.

**ACKNOWLEDGEMENTS**

We thank Chris Deloye, Natalia Ivanova, and Fred Rasio for very useful discussions. The data analysis was supported by the CXC CIAO software and CALDB. We have used the NASA NED and ADS facilities, and have extracted archival data from the *Chandra* archives. This work was supported by the *Chandra* GO grant G06-7079A (PI: Fabbiano) and sub-contract G06-7079B (PI: Kalogera). We acknowledge partial support from NASA contract NAS8-03060 (CXC). D.-W. Kim acknowledges support from *Chandra* archival research grant AR6-7008X, and A. Zezas from NASA LTSA grant NAG5-13056. T. Fragos acknowledges support by a Northwestern University Presidential Fellowship. G. Sivakoff, C. Sarazin, and A. Juett were supported in part by Hubble Grants HST-GO-10597.03-A, HST-GO-10582.02-A, and HST-GO-10835.01-A, and Chandra Grants GO7-8078X, GO7-8089A, and GO8-9085X. Fabbiano, Fragos, and Kalogera also acknowledge support by the Kavli Institute for Theoretical Physics at UCSB where part of this work was completed. This research was supported in part by the National Science Foundation under Grant PHY05-51164

**Figure Captions**

Figure 1. (top panels) Chandra X-ray images of three target galaxies. The red circles indicate X-ray sources and the green ellipses indicate the optical sizes of the galaxies ($D_{25}$). (bottom panels) HST optical images of the three galaxies. The blue crosses indicate globular clusters and the green ellipses indicate the optical sizes of the galaxies ($D_{25}$). The HST WFPC2 field-of-view is also marked for NGC 3379 and NGC 4278.

Figure 2. Histograms of GC X-ray counts (lower panels) and background-subtracted net counts (upper panels) for the GCs included in the stacking. The vertical lines in the upper panels are the medians of the distributions.

Figure 3. Cumulative luminosity distributions (in $L_X$) of detected field (solid) and GC (dashed) LMXBs. Both distributions are from the joint HST and Chandra field of view. Incompleteness is not corrected in this plot. The last bin with the arrow represents the contribution of undetected GC-LMXB from our stacking experiment (see section 5).

Figure 4. The X-ray luminosity function of field LMXBs (left) and GC-LMXBs (right) in the form of $dN/d\ln L_X$ against $L_X$. A single power-law + Gaussian model for field LMXBs and a broken power-law model for GC-LMXBs are overlaid (in red histograms) to illustrate different XLF shapes (see text for details). The vertical dotted line indicates the location of the bump (left) or the break luminosity (right). Also plotted in the bottom panels are sigma = (data – model) / error.

Figure 5. The X-ray luminosity function of LMXBs found inside $D_{25}$ (but excluding the central region, r < 10″) from all three galaxies. We fit the XLF with (a) a single power-law model, (b) a broken power-law model, and (c) a single power-law + Gaussian model. The blue histogram in (c) indicates the possible bump which consists of 15% of total LMXBs in number. Also plotted in the bottom panels are sigma = (data – model) / error. (d) The theoretical prediction taken from Fragos et al. (2008) is shown for comparison.

Figure 6. XLFs of individual galaxies: all LMXBs (left); field-LMXBs (middle); and GC-LMXBs (right). The three green vertical lines indicate 90%, 50% and 10% detection limits (from right to left) in each galaxy within the $D_{25}$ ellipse. Two diagonal lines with a slope of 1 (or β=2 in the differential XLF form) are drawn for visibility. For illustration purpose, we over-plot a single power-law model (left panels) and a single power-law with a Gaussian component (middle panels) and a broken power-law model (right panels).

Figure 7. (a) The number and (b) luminosity of LMXBs are plotted against GC specific frequency $S_N$ for NGC 3379, NGC 4697 and NGC 4278 (from left to right). The number and luminosity were determined within the $D_{25}$ ellipse, but excluding the central 10″ region. GC and field LMXBs are marked by red and blue colors, respectively. The dashed line indicates the best fit linear relation between $N(\text{LMXB})/L_K$ and $S_N$ for GC-LMXBs.

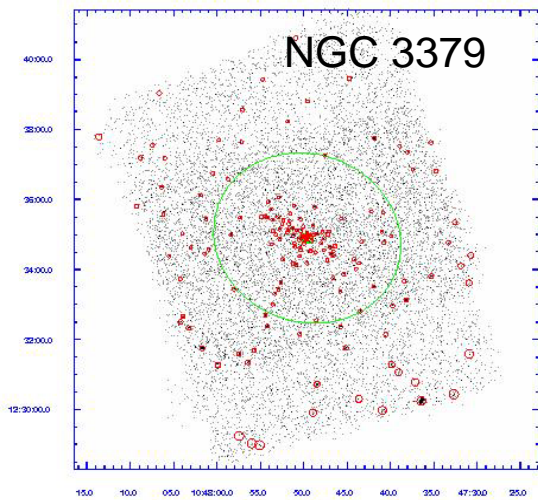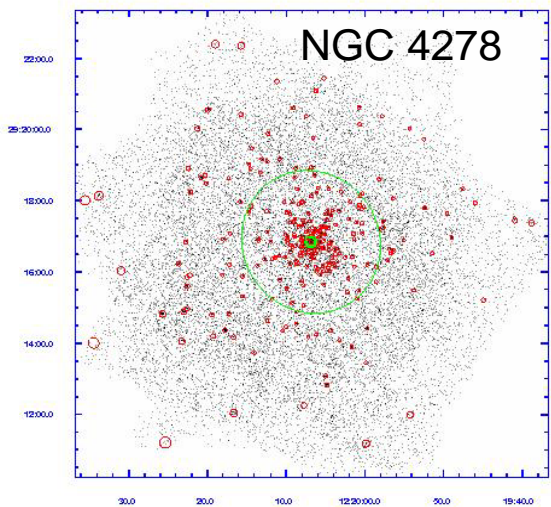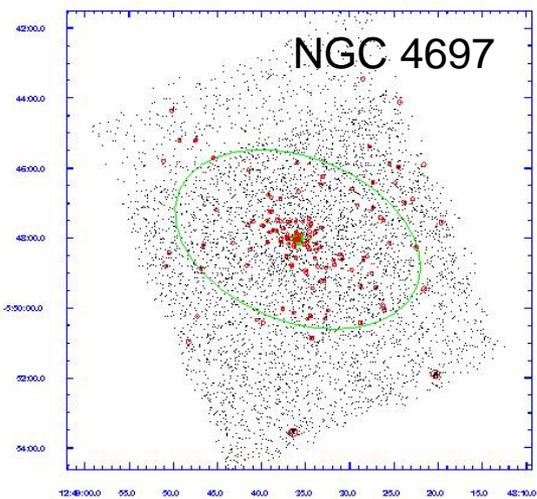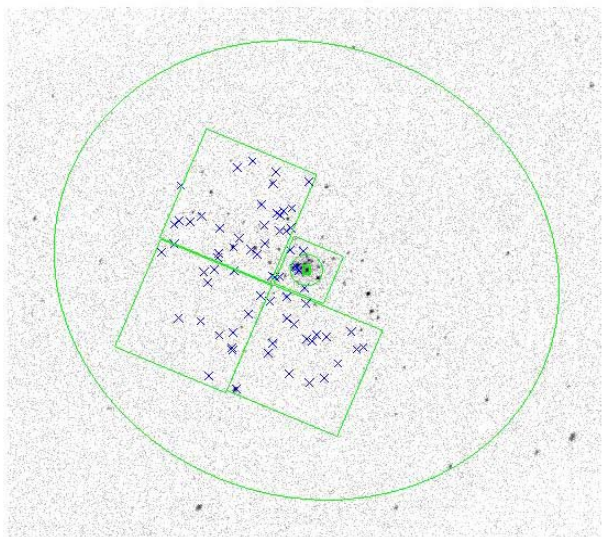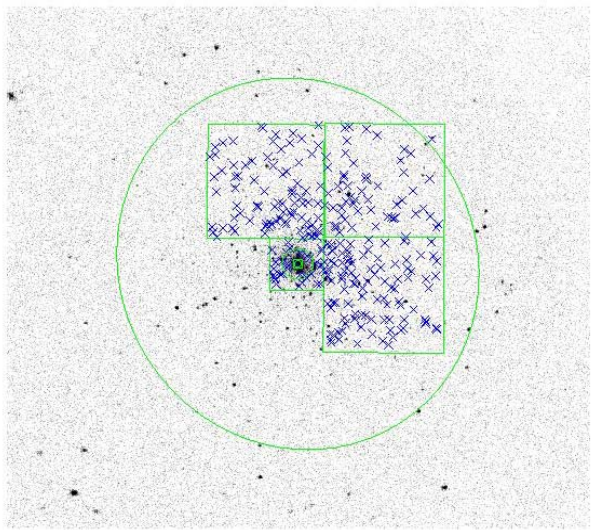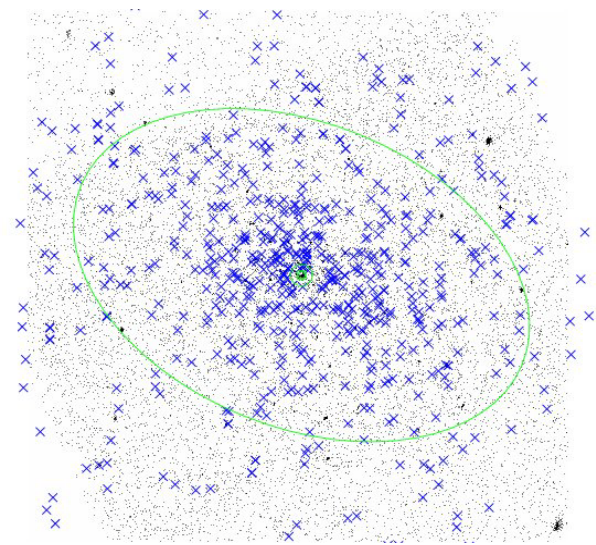

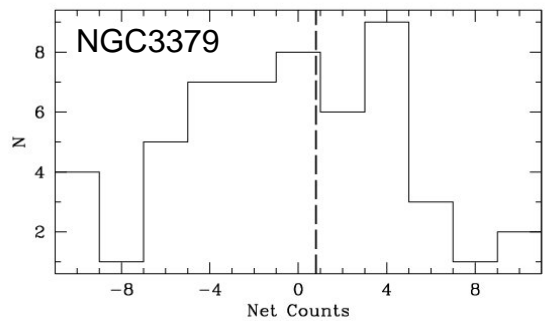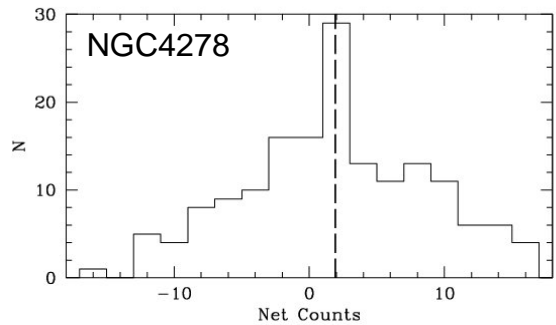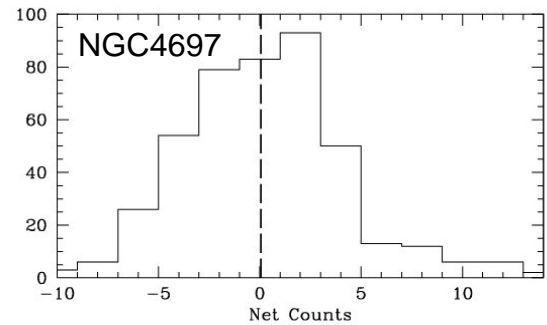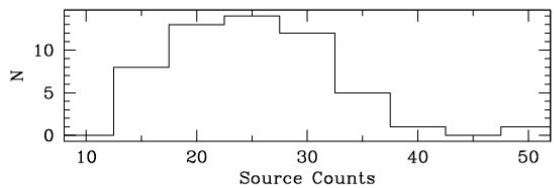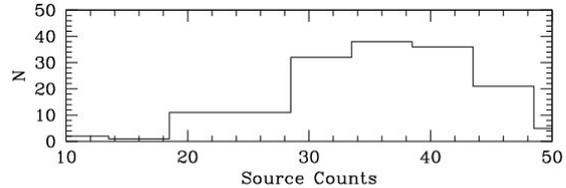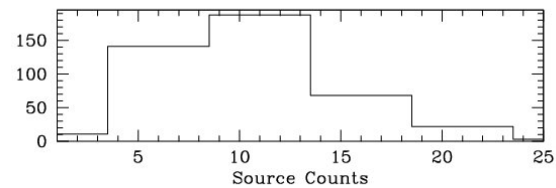

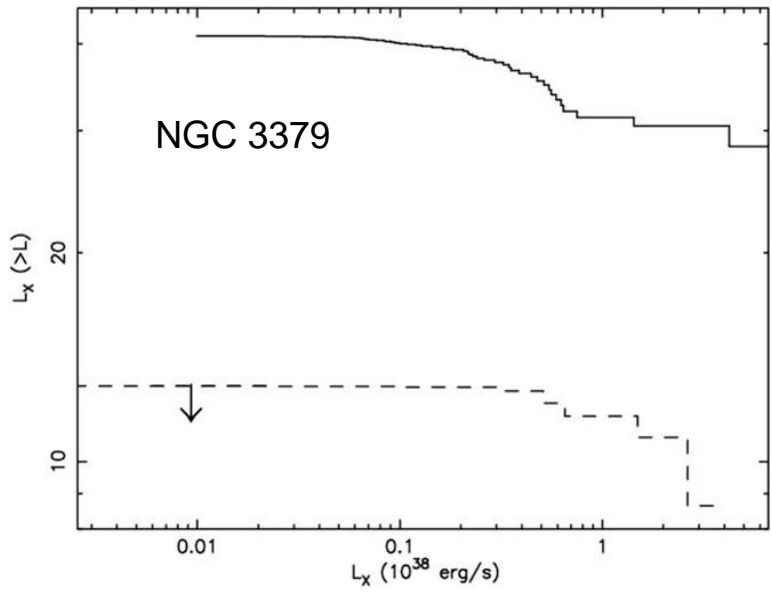
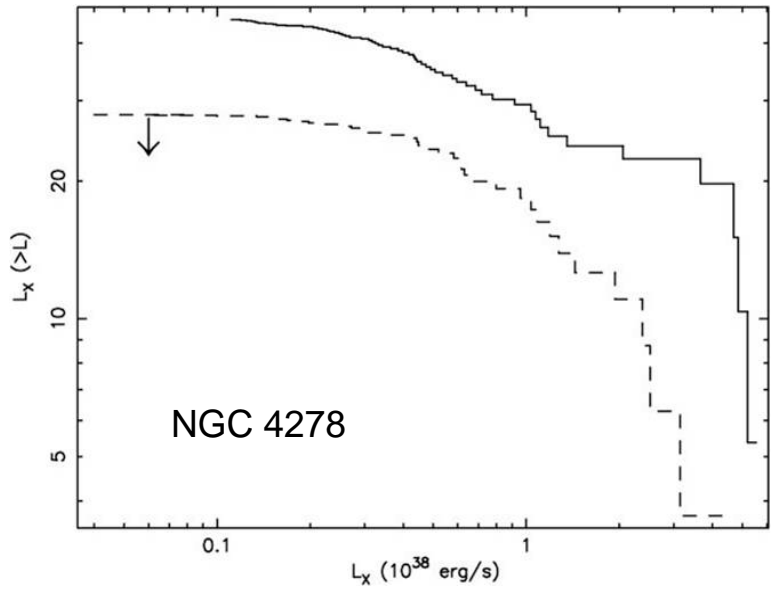
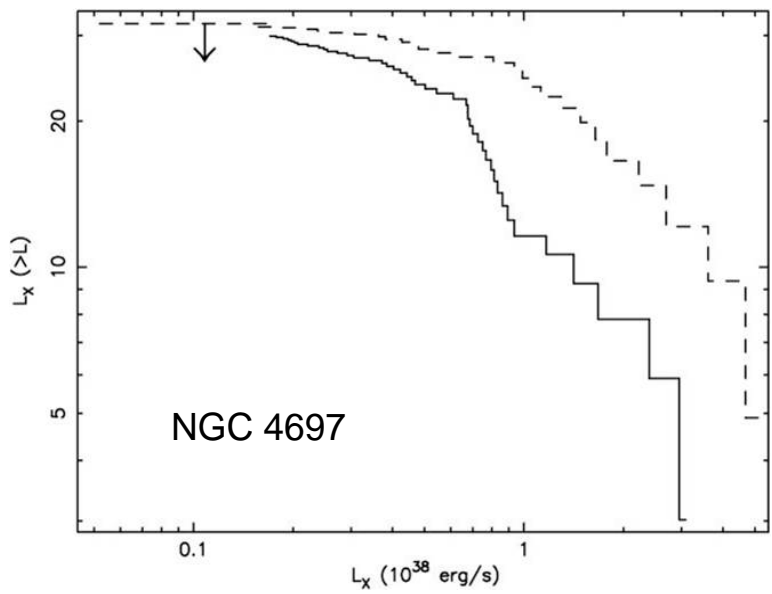

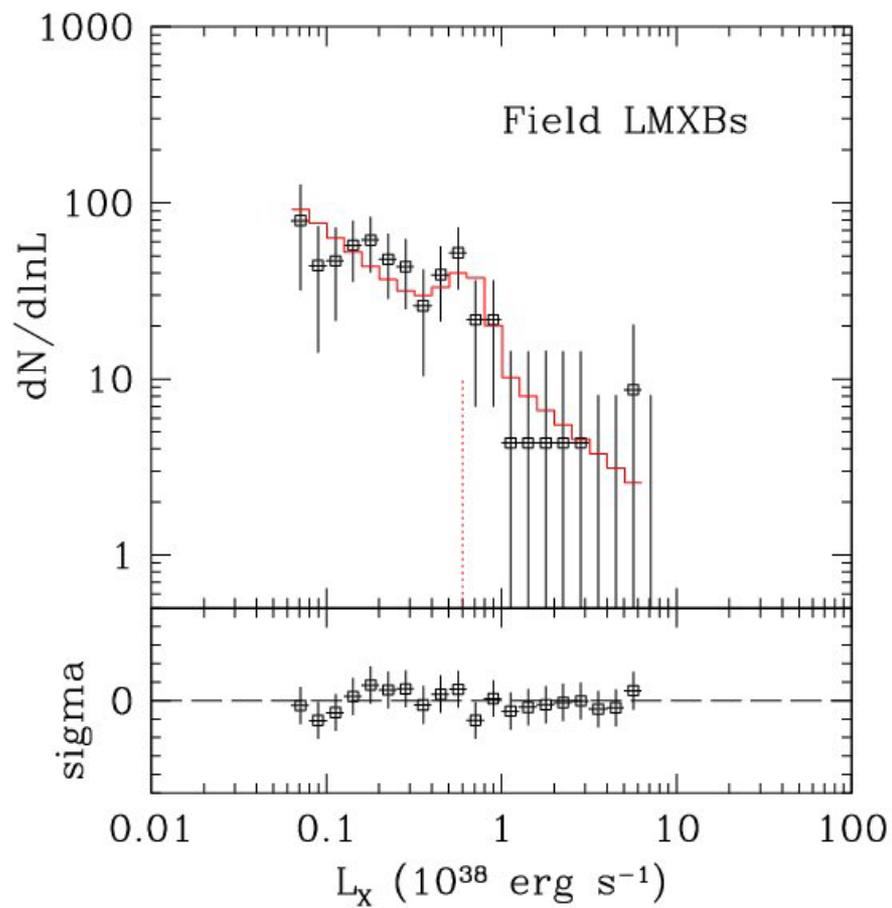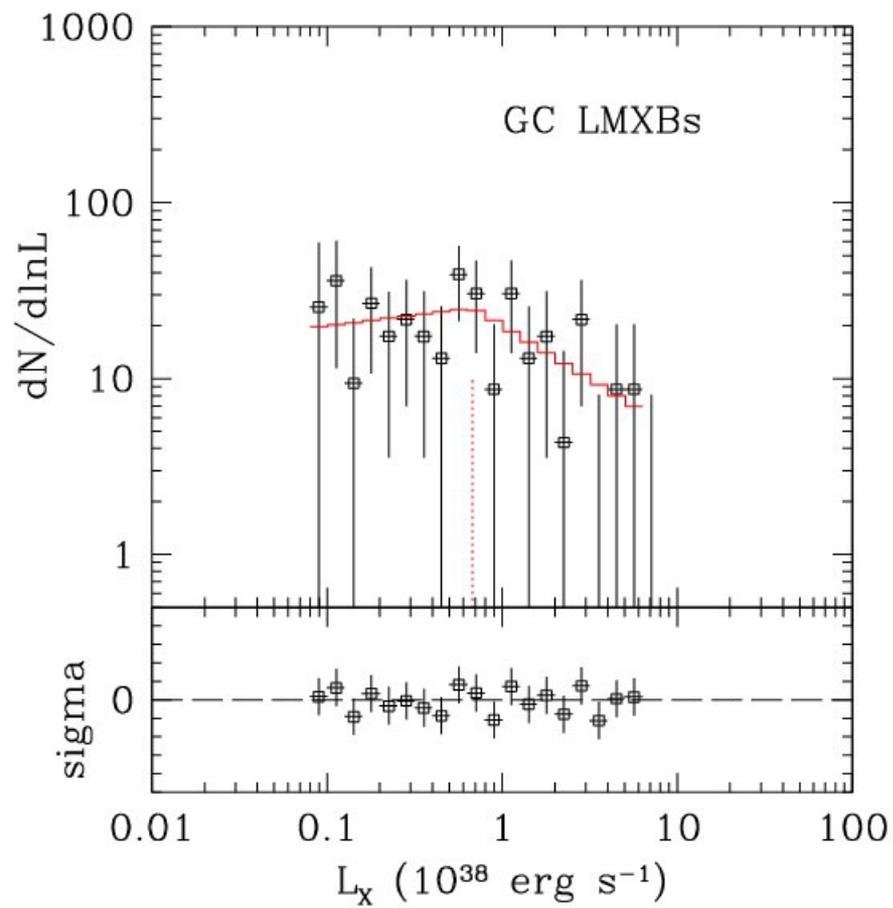

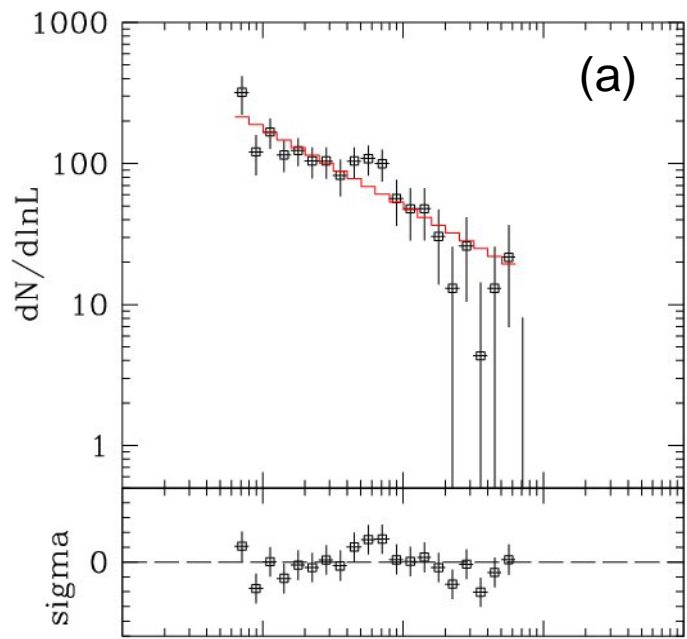
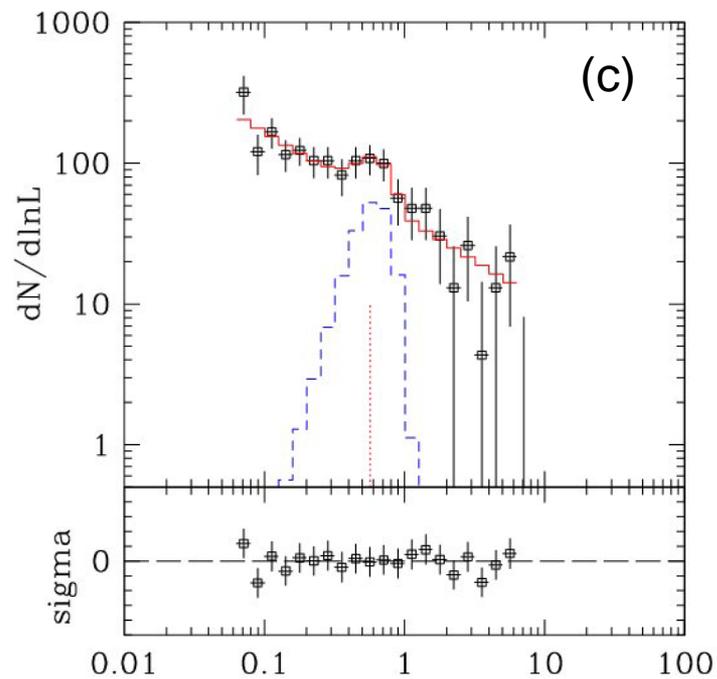
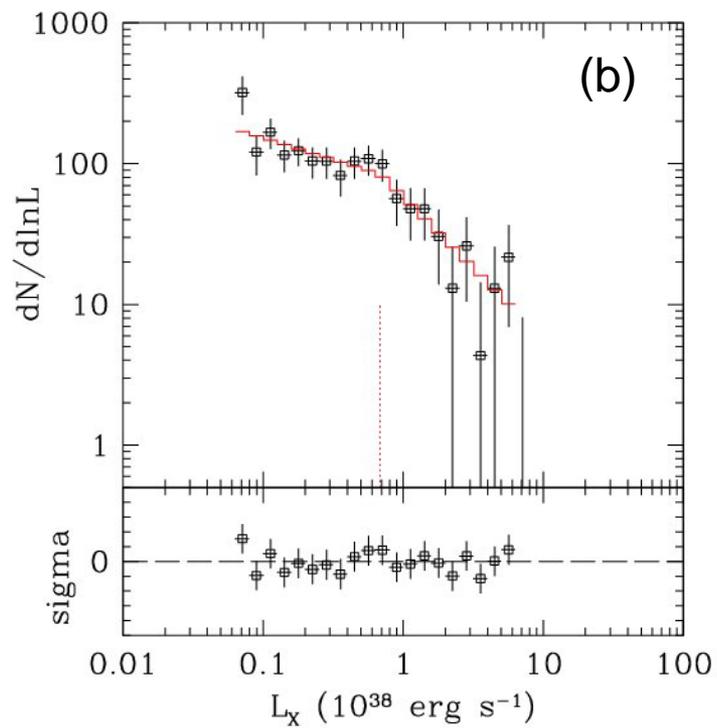
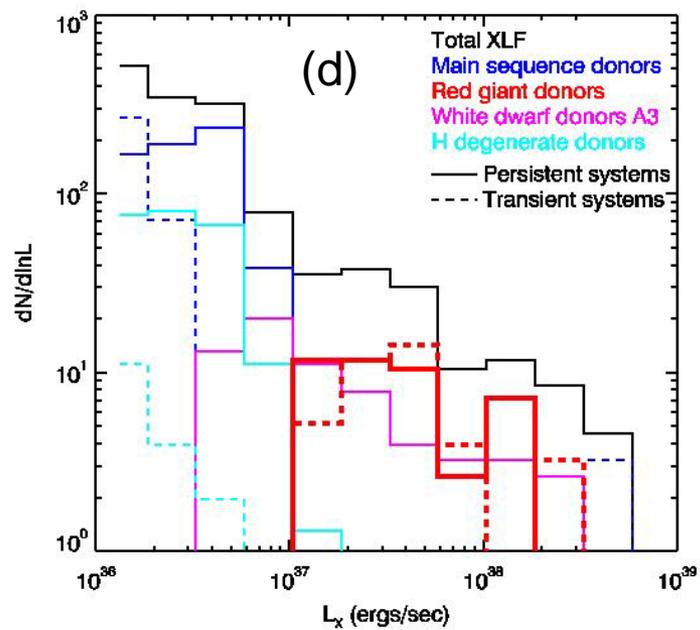

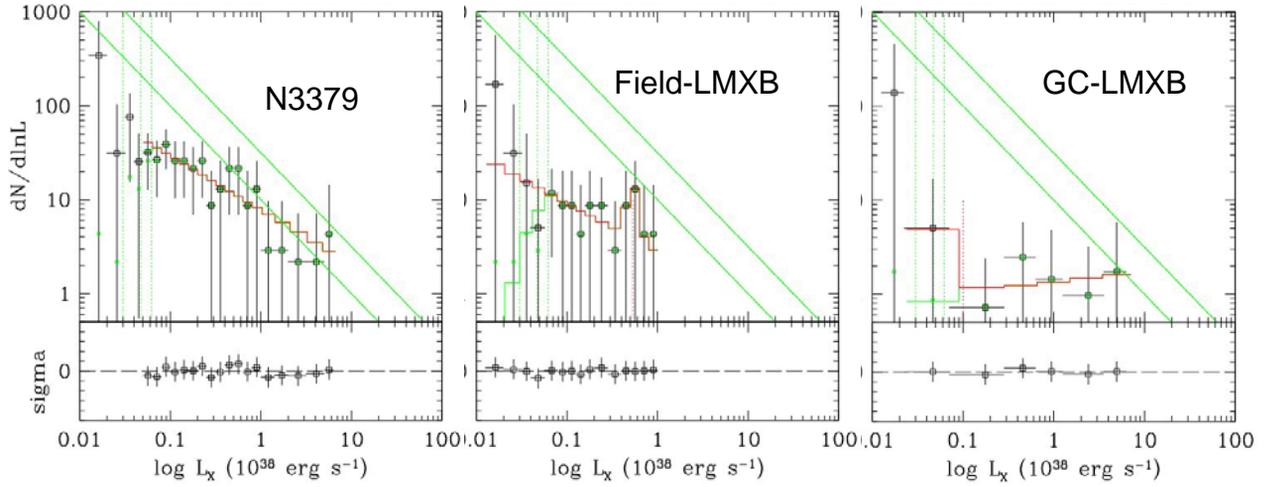
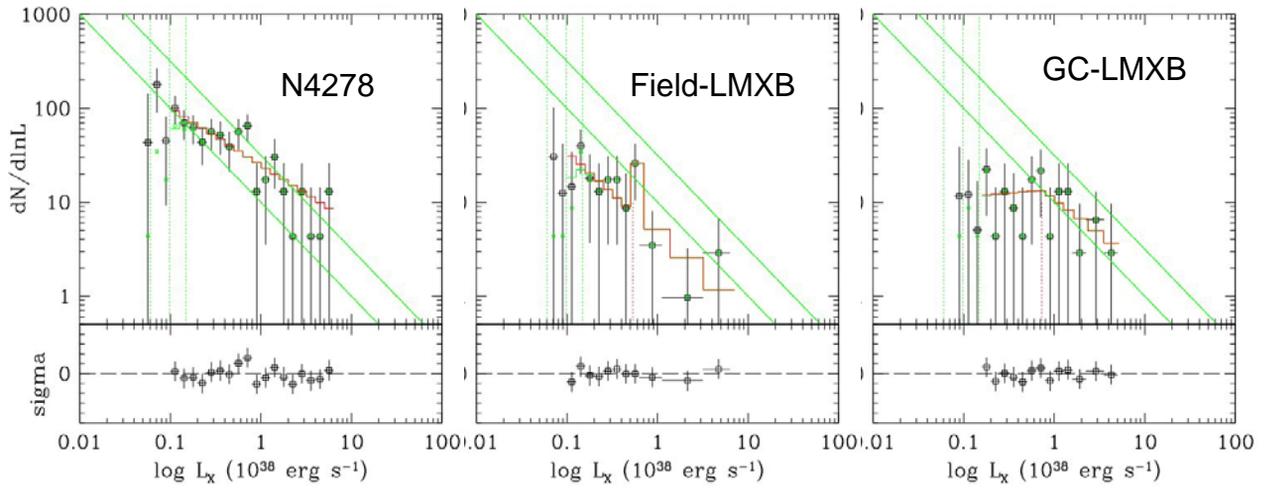
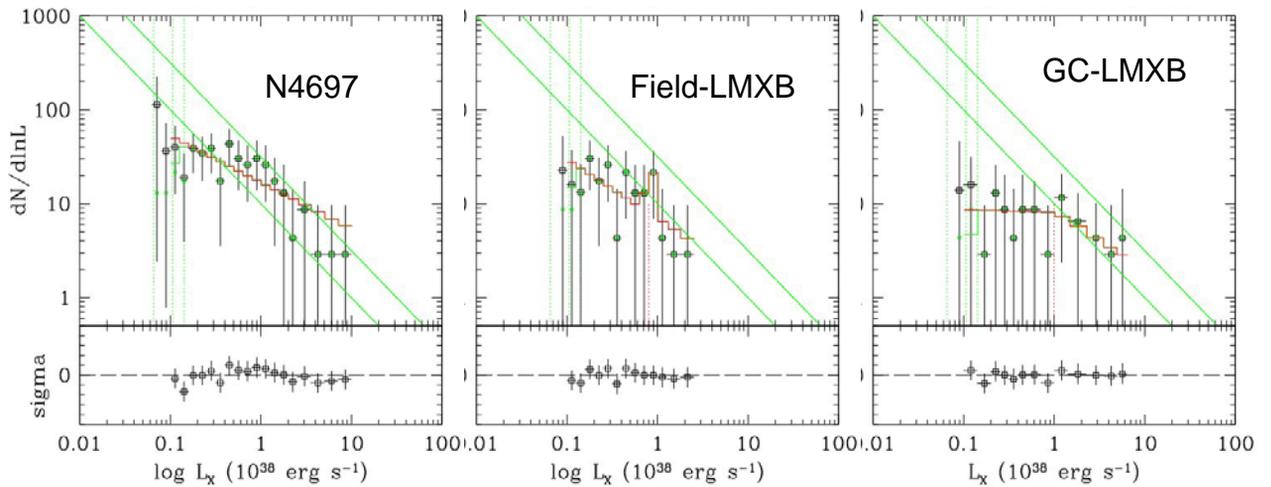

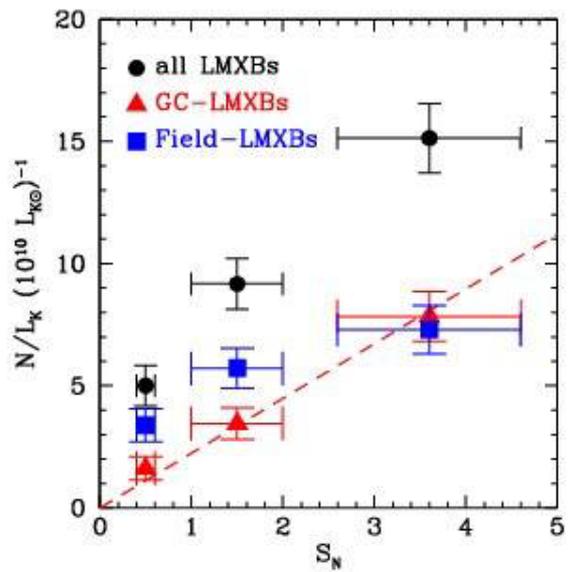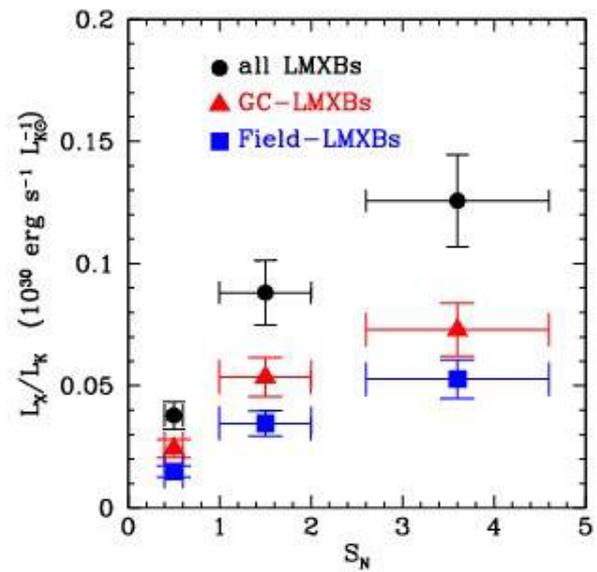